\documentclass[aps,twocolumn,showpacs]{revtex4}
\usepackage{amsmath,color,ulem}
\usepackage{multirow}
\usepackage{graphicx}
\usepackage{graphicx,color,dcolumn,booktabs,bm,multirow}
\usepackage{longtable,lscape}
\begin{document}

\title{Search for doubly charmed dibaryons in baryon-baryon scattering}

\author{Yao Cui$^1$}
\author{Yuheng Wu$^2$}
\author{Xinmei Zhu$^3$}\email{xmzhu@yzu.edu.cn}
\author{Hongxia Huang$^1$}\email{hxhuang@njnu.edu.cn}
\author{Jialun Ping$^1$}
\affiliation{$^1$School of Physics and Technology, Nanjing Normal University, Nanjing 210097, People's Republic of China}
\affiliation{$^2$Department of Physics, Yancheng Institute of Technology, Yancheng 224000, People's Republic of China}
\affiliation{$^3$Department of Physics, Yangzhou University, Yangzhou 225009, People's Republic of China}

\begin{abstract}
 We perform a systematical investigation of the doubly charmed dibaryon system with quantum numbers $IJ=01$, and strangeness numbers $S=0$, $-2$ and $-4$ in the framework of the chiral quark model. Two resonance states with strangeness numbers $S=-2$ is obtained in the $\Lambda\Omega_{cc}$ scattering channel, which are $\Xi_{cc}^{\ast}\Xi$ with resonance mass 5081 MeV and decay width 0.3 MeV, and the $\Xi_{c}\Xi_{c}^{\ast}$ state with the mass 5213 MeV and decay width 19.8 MeV, respectively. These two predicted charmed dibaryon candidates are worth searching for experimentally. Besides, we would like to emphasize that the multi-channel coupling calculation is important to confirm the existence of multiquark states. The coupling can shift the energy of the resonance, give the width to the resonance and even destroy the resonance. Therefore, to provide the necessary information for experiments to search for exotic hadron states, the coupling calculation between the bound channels and open channels is indispensable.
 \end{abstract}

\maketitle

\setcounter{totalnumber}{5}

\section{\label{sec:introduction}Introduction}

 Exploring the interactions between hadrons and the quest for exotic hadron states represent pivotal areas of inquiry in hadron physics. The study of hadron spectroscopy is a hot topic in particle physics research, with the study of double-heavy baryons being an important branch within this field. A long-standing challenge in this field has been the pursuit of dibaryons, with the notable discovery of the deuteron in 1932 marking a significant milestone ~\cite{deutron}. In 2014, the Wide Angle Shower Apparatus (WASA) detector at the Cooler Synchrotron (COSY)~\cite{WASA1, WASA2} collaboration established the narrow resonance state $d^{\ast}$ with $I(J^{P})=0(3^{+})$, and given the first clear-cut experimental evidence for the existence of a true dibaryon resonance~\cite{WASA3}. The $d^{\ast}$ (2380) could potentially be interpreted as either a $\Delta\Delta$ dibaryon state or a six-quark configuration, sparking extensive theoretical investigations across various theoretical approaches ~\cite{dstar1, dstar2, dstar3, dstar4}.

For the strange dibaryon, the progress of the $N\Omega$ searches in the experiment attracted more and more attention for this state, which was observed in Au+Au collisions by STAR experiment at the Relativistic Heavy Ion Collider (RHIC)~\cite{nomega1}. And before that, the dibaryon $N\Omega$ was investigated by different theoretical methods such as quark models~\cite{nomega2, nomega3, nomega4, nomega5, nomega6}, and the lattice QCD~\cite{nomega7, nomega8}. With the discovery of baryons with heavy quarks~\cite{CLEO:1994oxm, ARGUS:1993vtm, CLEO:2000ibb, CLEO:1999msf}, especially the exploration of charmed dibaryons has been significantly inspired by the experimental detection of the doubly charmed baryon, denoted as $\Xi_{cc}$, by the Large Hadron Collider beauty (LHCb) Collaboration~\cite{LHCb}. This observation has opened new avenues for understanding the behavior of hadrons  with heavy quarks.

For the heavy quark-containing dibaryons, the $N\Lambda_{c}$ system, which features a single heavy quark, has been examined from both the hadron level~\cite{c1} and the  quark level~\cite{c2}. The dibaryon systems with two heavy quarks were researched in the one-pion-exchange model~\cite{cc1}, which indicates that  a molecular bound state of two $\Lambda_{c}$'s is possible, where the tensor force plays a crucial role, although the binding energies are sensitive to the cutoff parameter. In Ref.~\cite{cc2}, the authors conduct a comprehensive investigation into the potential formation of loosely bound states, which are composed of either two charmed baryons or a charmed baryon and its antiparticle. This study is undertaken within the theoretical framework of the one-boson-exchange model. The numerical outcomes of their research suggest the non-existence of the H-dibaryon-like state, specifically the $\Lambda_{c}\Lambda_{c}$, and the possibility of the existence of four loosely bound deuteron-like states for the $\Xi_{c}\Xi_{c}$ and $\Xi_{c}^{\prime}\Xi_{c}^{\prime}$ systems with small binding energies and large rms radii. Within the framework of the one-pion-exchange model~\cite{cc3}, the researchers investigate the possible molecules composed of two heavy flavor baryons such as "$A_{Q}B_{Q}$" (Q=b, c), and come to a conclusion the long-range $\pi$ exchange force is strong enough to form molecules. $N\Xi_{cc}$, $N\Xi_{cc}^{\ast}$ and $\Sigma_{c}\Sigma_{c}^{\ast}$ were obtained by Ref.~\cite{D} within the framework of the quark delocalization color screening model (QDCSM). Zhi-Gang Wang and colleagues have identified indications of $\Sigma_{c}\Sigma_{c}$ and $\Sigma_{c}\bar{\Sigma}_{c}$ states with varying quantum numbers through the application of Quantum Chromodynamics (QCD) sum rules~\cite{Wang:2021pua}. Xiang Liu $et$ $al$.~\cite{Liu:2022rzu}, leveraging distinct symmetry principles, have delved into the study of doubly charmed dibaryon states characterized by the (qqqqcc) configuration. Their analysis suggests the potential existence of narrow, or even stable, states that are impervious to decay via strong interaction processes. Furthermore, numerous models have also been employed to investigate this system, including the string theory~\cite{Andreev:2024orz},  Bethe-Salpeter equation~\cite{Qi:2024dqz}, chiral effective field theory~\cite{Wang:2024riu} and  lattice QCD~\cite{Junnarkar:2022yak, Junnarkar:2024kwd}.


Besides, the dibaryon systems with three heavy quarks were also investigated from the lattice QCD~\cite{ccc1}, the QCD sum rule~\cite{ccc2}, one-boson-exchange~\cite{ccc3, ccc5} and the quark model~\cite{ccc4}. In Refs.~\cite{Meng:2017fwb, Yang:2019rgw}, the deuteron-like states composed of two doubly charmed baryons $\Xi_{cc}\Xi_{cc}$ and $\Xi_{cc}\bar{\Xi}_{cc}$ were systematically studied within the one-boson-exchange model. In addition, research on fully heavy dibaryon states has been emerging consistently~\cite{Lyu:2021qsh, Liu:2021pdu, Richard:2020zxb, full}.

 Quantum chromodynamics (QCD) is a theory describing strong interactions based on regular field theory. The equivalent degrees of freedom are quarks and gluons, and QCD is asymptotically free at high energies and can be solved precisely by perturbation theory. Generally, hadronic structure and hadron interactions belong to the low-energy physics of QCD, which are much harder to calculate directly from QCD because of the nonperturbative nature of QCD. One must rely on effective theories or models inspired by QCD to gain insight into the phenomena of the hadronic world. The constituent quark model is one of them, which transforms the complicated interactions between current quarks into dynamic properties of constituent quarks. The chiral quark model (ChQM) is a typical one of the constituent quark model. The ChQM was successfully used to calculate mesons~\cite{ChQM1, meson}, baryons, tetraquarks~\cite{4, chqm2}, pentaquarks~\cite{5} and dibaryons~\cite{nomega5}. In particular, for dibaryon systems, the ChQM is able to calculate the dibaryon systems from light to heavy quarks very well, such as nucleon-nucleon interaction~\cite{nn}, $N\Omega$~\cite{nomega5} and the fully heavy dibaryon systems~\cite{full}, which is consistent with the results of the lattice QCD.

Before undertaking this work, we had already investigated dibaryon states containing a single charm quark, predicting the possible existence of singly charmed dibaryon states~\cite{Cui:2024zfs}. Building on this foundation, the main focus of this paper is to study dibaryon states containing two charm quarks, providing theoretical support for the experimental search for double-charmed states. In the present work, we systematically investigate the doubly charmed dibaryons in the ChQM, where the effective potential between two baryons are evaluated, and the search of possible bound states are performed with the coupled channel effects. Moreover, based on the conservation of the quantum numbers and the limitation of the phase space, we also study the baryon-baryon scattering process to look for the existence of any resonance states in the doubly charmed dibaryon systems.

The structure of this paper is as follows. A brief introduction
of the quark model and calculation methods are given in Section II. Section III is devoted
to the numerical results and discussions. Section \uppercase\expandafter{\romannumeral4} is a
summary and the last section is Appendix, which shows the way of constructing wave functions.

\section{Quark model and calculation methods}
Phenomenological model is an important tool to analyze the nature of multi-quark states. Here, the chiral quark model(ChQM) is used to study the singly charmed dibaryon systems with $IJ=01$. In addition, the six-body problem is transformed into a two-body problem by using the resonance group method(RGM) for simplified calculations.

\subsection{The chiral quark model}
The model has become one of the most common approaches to describe hadron spectra, hadron-hadron interactions and multiquark states~\cite{ChQM(RPP)}. The construction of the ChQM is based on the breaking of chiral symmetry dynamics~\cite{chqm1, chqm2}. The model mainly uses one-gluon-exchange potential to describe the short-range interactions, a $\sigma$ meson exchange (only between u, d quarks) potential to provide the mid-range attractions, and Goldstone boson exchange potential for the long-range effects~\cite{chqm3}. In addition to the Goldstone bosons exchange, there are additional $D$ meson that can be exchanged between u/d and c quarks, $D_{s}$  meson that can be exchanged between s and c quarks, and $\eta_{c}$ that can be exchanged between any two quarks of the u, d, s and c quarks. In order to incorporate the charm quark well and study the effect of the $D$, $D_{s}, \eta_{c}$ meson exchange interaction, we extend the model to $SU(4)$, and add the interaction of these heavy mesons interactions. The extension is made in the spirit of the phenomenological approach of Refs. ~\cite{Glozman, Stancu}.
The detail of ChQM used in the present work can be found in the references~\cite{ChQM1,ChQM2,ChQM3}. In the following, only the Hamiltonian and parameters are given.

\begin{widetext}
\begin{eqnarray}
H &=& \sum_{i=1}^6 \left(m_i+\frac{p_i^2}{2m_i}\right) -T_c
+\sum_{i<j} \left[ V^{CON}(r_{ij})+V^{OGE}(r_{ij}) + V^{\sigma}(r_{ij})+ V^{OBE}(r_{ij})
\right], \\
V^{CON}(r_{ij})&=& -a_c {\boldsymbol \lambda}_i \cdot {\boldsymbol
\lambda}_j [r_{ij}^{2}+V_0],  \\
V^{OGE}(r_{ij})&=& \frac{1}{4}\alpha_{s} \boldsymbol{\lambda}_i \cdot
\boldsymbol{\lambda}_j
\left[ \frac{1}{r_{ij}}-\frac{\pi}{2}\left(\frac{1}{m_{i}^{2}}
 +\frac{1}{m_{j}^{2}}+\frac{4\boldsymbol{\sigma}_i\cdot
 \boldsymbol{\sigma}_j}{3m_{i}m_{j}}  \right)
 \delta(r_{ij})-\frac{3}{4m_{i}m_{j}r^3_{ij}}S_{ij}\right],
\end{eqnarray}
\begin{eqnarray}
 V^{\sigma } (r_{ij} )&=&-\frac{g_{ch}^{2} }{4\pi } \frac{\Lambda _{\sigma }^{2}m_{\sigma  }  }
 {\Lambda _{\sigma}^{2}-m_{\sigma }^{2} }\left [ Y\left ( m_{\sigma }r_{ij} \right )
  -\frac{\Lambda _{\sigma}}{m_{\sigma }}Y\left ( \Lambda _{\sigma }r_{ij} \right )\right ] \\
V^{OBE}(r_{ij}) & = & v^{\pi}(r_{ij}) \sum_{a=1}^3\boldsymbol{\lambda}_{i}^{a}\cdot
\boldsymbol{\lambda}_{j}^{a}+v^{K}(r_{ij})\sum_{a=4}^7\boldsymbol{\lambda}_{i}^{a}\cdot
\boldsymbol{\lambda} _{j}^{a} +v^{\eta}(r_{ij})\left[\left(\boldsymbol{\lambda} _{i}^{8}\cdot
\boldsymbol{\lambda} _{j}^{8}\right)\cos\theta_P-(\boldsymbol{\lambda} _{i}^{0}\cdot
\boldsymbol{\lambda}_{j}^{0}) \sin\theta_P\right] \nonumber \\
&& +v^{D}(r_{ij}) \sum_{a=9}^{12}
\boldsymbol{\lambda}_{i}^{a}\cdot \boldsymbol{\lambda}_{j}^{a}+v^{D_{s}}(r_{ij}) \sum_{a=13}^{14}
\boldsymbol{\lambda}_{i}^{a}\cdot \boldsymbol{\lambda}_{j}^{a}
+v^{\eta_{c}}(r_{ij})
\boldsymbol{\lambda}_{i}^{15}\cdot \boldsymbol{\lambda}_{j}^{15}, \\
v^{\chi}(r_{ij})&=& -\frac{g_{ch}^{2} }{4\pi }\frac{m_{\chi }^{2} }{12m_{i}m_{j}}\frac{\Lambda^2}{\Lambda^2-m_{\chi}^2}m_\chi \left\{ \left[Y(m_\chi r_{ij})-\frac{\Lambda^3}{m_{\chi}^3}Y(\Lambda r_{ij})
 \right] \boldsymbol{\sigma}_i \cdot \boldsymbol{\sigma}_j  \right. \nonumber \\
&& \left.+ \left[ H(m_\chi r_{ij})-\frac{\Lambda^3}{m_\chi^3}
 H(\Lambda r_{ij}) \right] S_{ij} \right\} \boldsymbol{\lambda}^F_i \cdot
 \boldsymbol{\lambda}^F_j, ~~~\chi=\pi,K,\eta,D,D_{s},\eta_{c},  \\
 S_{ij} & = &  \frac{({\boldsymbol \sigma}_i \cdot {\boldsymbol r}_{ij})
({\boldsymbol \sigma}_j \cdot {\boldsymbol
 r}_{ij})}{r_{ij}^2}-\frac{1}{3}~{\boldsymbol \sigma}_i \cdot {\boldsymbol
\sigma}_j.
\label{H}
\end{eqnarray}
\end{widetext}
 Where $T_c$ is the kinetic energy of the center of mass; $S_{ij}$ is quark tensor operator. We only consider
 the $S-$wave systems at present, so the tensor force dose not work here; $Y(x)$ and $H(x)$ are
 standard Yukawa functions~\cite{ChQM(RPP)}; $\alpha_{ch}$ is the chiral coupling constant, determined as usual from the $\pi$-nucleon coupling constant; $\alpha_{s}$ is the quark-gluon coupling constant~\cite{ChQM1}. Here $m_{\chi}$ is the mass of the mesons, which are experimental value; $\Lambda_{\chi}$ is the cut-off parameters of different mesons, which can refer to Ref~\cite{D}. The coupling constant $g_{ch}$ for scalar chiral field is determined from the $NN\pi$ coupling constant through
 \begin{equation}
 \frac{g_{ch}^{2} }{4\pi } =(\frac{3}{5} )^{2} \frac{g_{\pi NN}^{2} }{4\pi}\frac{m_{u,d}^{2} }{m_{N}^{2} }
 \end{equation}
All other symbols have their usual meanings.
\begin{table}[ht]
\renewcommand{\arraystretch}{1.5}
\begin{center}
\caption{Model parameters} 

\begin{tabular}{lccccc} \hline \hline
& ~~~~~~$b$~~~~ & ~~~~$m_{u,d}$~~~~ & ~~~~$m_{s}$~~~~ & ~~~~$m_{c}$~~~~ & ~~~~$m_{b}$~~~~   \\
& (fm) & (MeV) & (MeV) & (MeV) & (MeV)   \\ \hline\noalign{\smallskip}
ChQM & 0.52088  & 313  &  590 & 1700 &    5105   \\ \hline\noalign{\smallskip}
 & $ a_c$ & $V_{0}$ &  $\alpha _{s_{qq} }  $ & $\alpha _{s_{qs} }  $ &  $\alpha _{s_{ss} } $   \\
 & (MeV\,fm$^{-2}$) & (fm$^{2}$) &     \\\hline\noalign{\smallskip}
ChQM & 49.350 & -1.0783  & 0.67321&   0.85644 &  0.71477  \\
\hline
 &  $\alpha _{s_{qc} }  $ & $\alpha _{s_{sc} }  $ &  $\alpha _{s_{cc} } $   \\ \hline
ChQM &0.59301  &0.60775&   1.0807\\\hline\hline
\end{tabular}
\label{parameters}
\end{center}
\end{table}

\begin{table}[ht]
\renewcommand{\arraystretch}{1.5}
\begin{center}
\caption{The calculated masses (in MeV) of the  baryons in ChQM.
 Experimental values are taken from the Particle Data Group (PDG)~\cite{PDG}. }
\begin{tabular}{lcccccc} \hline \hline
               & ~~$N$~~  & ~~$\Delta$~~  & ~~$\Lambda$~~   & ~~$\Sigma$~~ & ~~$\Sigma^{*}$~~   & ~~$\Omega$~~  \\ \hline
ChQM          & 933   & 1254     & 1100  & 1201  & 1370  & 1664     \\
 Exp.          & 939  &1233      &1116    &1189   &1315    &1672   \\ \hline
                & ~~$\Xi$~~ &~~$\Xi^{*}$~~ &  ~~$\Lambda_{c}$~~  & ~~$\Sigma_{c}$~~ &   ~~$\Sigma^{*}_{c}$~~   & ~~$\Xi_{c}$~~ \\ \hline
 ChQM           & 1338  & 1507   &2225 &  2416 &  2449       & 2450 \\
 Exp.          &1385    &1530    &2286  & 2455  &  2520      & 2470 \\ \hline
                & ~~$\Xi^{'}_{c}$~~ & ~~$\Xi^{*}_{c}$~~&~~$\Xi_{cc}$~~   &~~$\Omega_{c}$~~  &~~$\Omega^{*}_{c}$~~   &\\ \hline
ChQM           & 2546  & 2571       & 3493    &2696  &2714   \\
 Exp.           & 2578       & 2645  & 3519     &2695   &2700     \\
\hline\hline
\end{tabular}
\label{baryons}
\end{center}
\end{table}

All parameters were determined by fitting the masses of the baryons of light and heavy flavors. The model parameters and the fitting masses of baryons are shown in Table~\ref{parameters} and Table~\ref{baryons}, respectively.

\subsection{Calculation methods}
In this work, RGM~\cite{RGM1} is used to carry out a dynamical calculation. In the framework of RGM, which split the dibaryon system into two clusters,
By expanding the relative motion wave function between two clusters in the RGM equation by gaussians, the integro-differential equation of RGM can be reduced to an algebraic equation, which is the generalized eigen-equation. Then by solving the eigen-equation, the energy of the system can be obtained. Besides, to keep the matrix dimension manageably small, the baryon-baryon separation is taken to be less than 6 fm in the calculation. The details of solving the RGM equation can be found in Ref~\cite{RGM1}.
the main feature of RGM is that for a system consisting of two clusters, it can assume that the two clusters are frozen inside, and only consider the relative motion between the two clusters

So the conventional ansatz for the two-cluster wave function is:
\begin{equation}
\psi_{6q} = {\cal A }\left[[\phi_{B_{1}}\phi_{B_{2}}]^{[\sigma]IS}\otimes\chi_{L}(\boldsymbol{R})\right]^{J}, \label{6q}
\end{equation}
where the symbol ${\cal A }$ is the anti-symmetrization operator. With the $SU(4)$ extension, both the light and heavy quarks are considered as identical particles. So we use ${\cal A } = 1-9P_{36}$. $[\sigma]=[222]$ gives the total color symmetry and all other symbols have their usual meanings. $\phi_{B_{i}}$ is the $3-$quark cluster wave function. From the variational principle, after variation with respect to the relative motion wave function $\chi(\boldsymbol{\mathbf{R}})=\sum_{L}\chi_{L}(\boldsymbol{\mathbf{R}})$, one obtains the RGM equation
\begin{equation}
\sum_{j} C_{j}H_{i,j}= E \sum_{j} C_{j}N_{i,j}.
\end{equation}
where $H_{i,j}$ and $N_{i,j}$ are the Hamiltonian matrix elements and overlaps, respectively. In our previous work~\cite{Cui:2024zfs}, we presented a detailed derivation of this process, and further details can be found in Ref.\cite{RGM1}. Besides, to keep the matrix dimension manageably small, the baryon-baryon separation is taken to be less than 6 fm in the calculation. By solving the generalized energy problem, we can obtain the energy and the corresponding wave functions of the dibaryon system. On the basis of RGM, we can further calculate scattering problems to find resonance states. Below, we will introduce the calculation process of scattering phase shifts from two aspects: single-channel and multi-channel coupling.

For the single-channel scattering problem, the relative wave function of the baryon-baryon is $\chi({{R}})$. First, we will utilize the Kohn-Hulth$\acute{e}$n-Kato(KHK) variational method~\cite{KHK} to simplify the equation-solving process.

Let $\chi({{R}})=u({R})/{R}$, we know that for a scattering wave function $u_{L}({R})$~ it should satisfy the following boundary conditions:
\begin{equation}
\left\{\begin{matrix}u_{L}(0)=0~~~~~~~~~~~~~~~~~~~~~~~~~~~~~~~~~~~~~~~~~~~~~~~~~~~ \\
u_{L}(R)=\left [ h_{L}^{(-)}(k,R) +S_{L}h_{L}^{(+)}(k,R)\right ]R,~R>R_{C}
\end{matrix}\right.
\end{equation}
Where $h_{L}^{\pm}$ are the $L$th spherical Hankel functions, $k$ is the momentum of the relative motion with $k =\sqrt{2\mu E_{cm} }$, $\mu$ is the reduced mass of two baryons of the open channel, $E_{cm}$ is the incident energy of the relevant open channels, and $R_{C}$ is a cutoff radius beyond which all of the strong interactions can be disregarded. The subscript $L$ indicates the relative motion wave function between two baryons. For ease of notation, in the following expressions ($u_t$, $u_i$, $u_i^{in}$, $c_i$ and $s_i$), the subscript will be omitted; $S_L$ represents the scattering matrix, and the relationship between the scattering matrix and the scattering phase shift is satisfied:
\begin{equation}\label{delta}
S_{L}=\left|S_{L}\right|e^{2i\delta_{L}}
\end{equation}
In order to obtain the wave function $u_L(R)$ and the scattering matrix $S_L$, we first introduce a trial wave function $u_t(R)$ that satisfies the boundary conditions mentioned above:
\begin{equation}
\left\{\begin{matrix}u_{t}(0)=0~~~~~~~~~~~~~~~~~~~~~~~~~~~~~~~~~~~~~~~~~~~~~~~~~~ \\
u_{t}(R)=\left [ h_{L}^{(-)}(k,R) +S_{t}h_{L}^{(+)}(k,R)\right ]R,~R>R_{C}
\end{matrix}\right.
\end{equation}
We expand this trial wave function using a series of known wave functions $u_i(R)$:
\begin{equation}\label{14}
u_{t}(R)=\sum_{i=0}^{n}c_iu_i(R)
\end{equation}
where $c_{i}$ are the expansion coefficients, and $c_{i}$ satisfy $\sum_{i=1}^{n}c_{i}=1$. n is the number of Gaussion bases (which is determined by the stability of the results). The function $u_i(R)$ satisfies the condition:
\begin{equation}
u_{i}\left ({R}\right )   \nonumber \\
\end{equation}
\begin{equation}\label{15}
=\left\{\begin{matrix}
& \alpha _{i}u_{i}^{(in)}\left(R \right ), ~~~~~~~~~~~~~~~~~~~~~~&{R}\le{R}_{C}  \\
& \left [ h_{L}^{(-)}\left ( {k},{R}\right )+s_{i}h_{L}^{(+)}\left ( {k},{R}\right )\right ]{R}, ~~~~~&{R}\ge {R}_{C}
\end{matrix}\right.
\end{equation}
Here, $u_i^{in}(R)$ is the relative motion wave function between the two heavy quarks used in solving the double-heavy baryon bound state problem, represented by Gaussian wave functions, with the specific form being:
\begin{equation}
\frac{u_{i}^{(in)}(R)}{R}=\sqrt{4\pi}\left ( \frac{3}{2\pi b^{2} }  \right )e^{-\frac{3}{4b^{2} }\left ({R}^{2}+{r_{i}}^{2}\right )}i^Lj_{L}\left ( -i\frac{3}{2b^{2} }Rr_{i}\right )
\end{equation}
Substituting Eq.~\ref{15} into Eq.~\ref{14}, we obtain:
\begin{equation}
u_{t}\left ({R}\right )   \nonumber \\
\end{equation}
\begin{equation}
=\left\{\begin{matrix}
& \sum_{i=0}^{n}c_i\alpha _{i}u_{i}^{(in)}\left(R \right ), ~~~~~~~~~~~~~~~~~~~~~~~~~~~~&{R}\le{R}_{C}  \\
& \sum_{i=0}^{n}\left [ c_i h_{L}^{(-)}\left ( {k},{R}\right )+c_i s_{i}h_{L}^{(+)}\left ( {k},{R}\right )\right ]{R}, ~&{R}\ge {R}_{C}
\end{matrix}\right.
\end{equation}
where
\begin{equation}
S_{t}=\sum_{i=0}^{n}c_{i}s_{i}
\end{equation}
Substitute into the $Schr\ddot{o}dinger$ projection equation:
\begin{align}
\left\langle \delta \Psi^{\prime} \right| H-E \left|\Psi \right\rangle & = 0
\end{align}
We arrive at:
\begin{equation}
\sum_{j=1}^{n} \mathcal{L}_{ij} c_{j} =\mathcal{M}_{i}~~~~~~~~(i=0,1,...,n)
\label{M}
\end{equation}
with
\begin{equation}
\mathcal{L}_{ij}=\mathcal{K}_{ij}-\mathcal{K}_{i0}-\mathcal{K}_{0j}+\mathcal{K}_{00}
\end{equation}
\begin{equation}
\mathcal{M}_{i}=\mathcal{M}_{ij}-\mathcal{K}_{i0}
\end{equation}
and
\begin{align}
\mathcal{K}_{ij}=&\left\langle {\phi_{A}}{\phi_{B}}\frac{u_{i}(R)}{R}\cdot Y_{L,M}(\hat{R})\left| H-E \right| \right.\nonumber \\
&\left.\cdot \mathcal{A}\left [ {\phi_{A}}{\phi_{B}}\frac{{u}_{j}(R)}{R}Y_{L,M}(\hat{R}) \right ] \right \rangle
\end{align}
By solving Eq.(\ref{M}) we obtain the expansion coefficients $c_{i}$. Then, the scattering matrix $S_{L}$ is given by
\begin{equation}
S_{L}=S_{t}-i \alpha\sum_{i=0}^{n}\mathcal{K}_{0i}c_{i}.
\end{equation}
where $\alpha=\frac{\mu}{\hbar k}$.

For the case of multi-channel coupling, the total wave function can be written as:
\begin{equation}
\Psi^{(c)}=\sum_{\gamma}\mathcal{A}[\Phi_{\gamma}\chi_{\gamma}^{(c)}(R_{\gamma })]+\Omega^{(c)}~~(c=\alpha,\beta)
\end{equation}
where $\Psi^{(c)}$ denotes the relative motion wave function of the incident channel (c) along with all the outgoing channels; $\gamma$ represents all the two-body channels. Here, $\Omega^{(c)}$ denotes the decay residue amplitude without the inclusion of the previous term, which is generally negligible and can usually be omitted. The asymptotic behavior of $\chi_{\gamma}^{(c)}(R_{\gamma})$ is:
\begin{align}
\chi_{\gamma}^{(c)}\left(R_{\gamma}\right)=&\chi_{\gamma}^{(-)}\left(k_{\gamma}, R_{\gamma}\right) \delta_{\gamma c}+S_{\gamma c} \chi_{\gamma}^{(+)}\left(k_{\gamma}, R_{\gamma}\right),\nonumber\\
 &~~~~~~~~~~~~~~~~~~~~~~~~~~~~~~~~~~R_{\gamma}>R_{\gamma}^{(c)}
\end{align}
Here, $S_{\gamma c}$ represents the S-matrix element of $c \to \gamma$; and $\chi_{\gamma}^{(\pm)}$ satisfies the condition:

(i)For open channels:
\begin{equation}
\chi_{\alpha}^{( \pm)}\left(k_{\alpha}, R_{\alpha}\right)=\frac{1}{\sqrt{v_{\alpha}}} h_{L_{\alpha}}^{( \pm)}\left(k_{\alpha}, R_{\alpha}\right), \quad R_{\alpha}>R_{\alpha}^{C}
\end{equation}

(ii)For closed channels:
\begin{equation}
\chi_{\alpha}^{( \pm)}\left(k_{\alpha}, R_{\alpha}\right)=W_{L_{\alpha}}^{( \pm)}\left(k_{\alpha}, R_{\alpha}\right), \quad R_{\alpha}>R_{\alpha}^{C}
\end{equation}

Here, $v_{\alpha}$ is the relative velocity of motion, $ v_{\alpha}=\hbar k_{\alpha} / \mu_{\alpha}$, $W_{L_{\alpha}}^{( \pm)}\left(k_{\alpha}, R_{\alpha}\right)$ have been discussed in Ref.\cite{KHK}

Similarly to the single-channel calculation, we first introduce a trial wave function $\Psi_t^{(c)}$:
\begin{equation}
\Psi_t^{(c)}=\sum_{\gamma}\mathcal{A}[\Phi_{\gamma}\chi_{\gamma, t}^{(c)}(R_{\gamma })]+\sum_{\nu}b_{\nu}^{(c)}\Omega_{\nu,t}~~(c=\alpha,\beta)
\end{equation}
The asymptotic behavior of $\chi_{\gamma, t}^{(c)}(R_{\gamma})$ is:
\begin{align}
\chi_{\gamma, t}^{(c)}\left(R_{\gamma}\right)=&\chi_{\gamma}^{(-)}\left(k_{\gamma}, R_{\gamma}\right) \delta_{\gamma c}+S_{\gamma c, t} \chi_{\gamma}^{(+)}\left(k_{\gamma}, R_{\gamma}\right),\nonumber\\
 &~~~~~~~~~~~~~~~~~~~~~~~~~~~~~~~~~~R_{\gamma}>R_{\gamma}^{(c)}
\end{align}
Here, we also expand $\chi_{\gamma}^{(c)}\left(k_{\gamma}, R_{\gamma}\right)$ using a series of known wave functions:
\begin{equation}
\chi_{\gamma, t}^{(c)}\left({R}_{\gamma}\right)=\sum_{i=0}^{n_{\gamma}} C_{\gamma i}^{(c)}\chi_{\gamma i}\left({R}_{\gamma}\right),~~~(c=\alpha, \beta)
\end{equation}
\begin{equation}
\chi_{\gamma i}\left({R}_{\gamma}\right)\\ \nonumber
\end{equation}
\begin{equation}
=\left\{\begin{matrix}
&\alpha_{\gamma i} u_{\gamma i}^{(in)}\left({R}_{\gamma}\right), ~~~~~~~~~~~~~~~~~~~~~~~~~~& {R}_{\gamma}<{R}_{\gamma}^{c} \\
&\left [ u_{\gamma}^{(-)}\left({k}_{\gamma}, {R}_{\gamma}\right)+s_{\gamma i} u_{\gamma}^{(+)}\left({k}_{\gamma}, {R}_{\gamma}\right) \right ]{R}_{\gamma} , & {R}_{\gamma}>{R}_{\gamma}^{c}
\end{matrix}\right.
\end{equation}
We can obtain:
\begin{align}\label{32}
\sum_{i=0}^{n_{\gamma}} C_{\gamma i}^{(c)}=&\delta_{\gamma c} \quad(c=\alpha, \beta) \\
\sum_{i=0}^{n_{\gamma}} C_{\gamma i}^{(c)} s_{\gamma i}=&S_{\gamma c, t} . \quad(c=\alpha, \beta)
\end{align}
Using the same method as in the single-channel calculation, we ultimately arrive at two sets of linear equations:
\begin{align}\label{33}
\sum_{\delta} \sum_{j=1}^{n_{\delta}} L_{\gamma i, \delta j} C_{\delta j}^{c}+\sum_{\mu} L_{\gamma i, \mu} b_{\mu}^{(c)}=&M_{\gamma i}^{(c)} \quad(c=\alpha, \beta)
\end{align}
\begin{align}\label{34}
\sum_{\delta} \sum_{j=1}^{n_{\delta}} L_{\nu, \delta_{j}} C_{\delta_{j}}^{(c)}+\sum_{\mu} L_{\nu, \mu} b_{\mu}^{(c)}=&M_{\nu}^{(c)} \quad(c=\alpha, \beta)
\end{align}
Eq.\ref{33} and \ref{34} are valid for all $\gamma$, and $ i=1\to n_{\gamma} $.

\begin{align}
\mathcal{K}_{\gamma i, \delta j} & =\int A_{\gamma}\left[\phi_{\gamma}^{\dagger} \chi_{\gamma i}\right](H-E) A_{\delta}\left[\phi_{\delta} \chi_{\delta j}\right] d \tau \\
\mathcal{K}_{\nu, \delta j} & =\int \tilde{\Omega}_{\bar{\nu}, t}^{\dagger}(H-E) A_{\delta}\left[\phi_{\delta} \chi_{\delta j}\right] d \tau \\
\mathcal{K}_{\gamma i, \mu} & =\int A_{\gamma}\left[\phi_{\gamma}^{\dagger} \chi_{\gamma i}\right](H-E) \Omega_{\mu, t} d \tau \\
\mathcal{K}_{\nu, \mu} & =\int \tilde{\Omega}_{\bar{\nu}, t}^{\dagger}(H-E) \Omega_{\mu, t} d \tau \\
\mathcal{L}_{\gamma i, \delta j} & =\mathcal{K}_{\gamma i, \delta j}-\mathcal{K}_{\gamma 0, \delta j}-\mathcal{K}_{\gamma i, \delta 0}+\mathcal{K}_{\gamma 0, \delta 0} \\
\mathcal{L}_{\nu, \delta j} & =\mathcal{K}_{\nu, \delta j}-\mathcal{K}_{\nu, \delta 0} \\
\mathcal{L}_{\gamma i, \mu} & =\mathcal{K}_{\gamma i, \mu}-\mathcal{K}_{\gamma 0, \mu} \\
\mathcal{L}_{\nu, \mu} & =\mathcal{K}_{\nu, \mu} \\
\mathcal{M}_{\gamma i}^{(c)} & =-\mathcal{K}_{\gamma i, c 0}+\mathcal{K}_{\gamma 0, c 0} \quad(c=\alpha, \beta) \\
\mathcal{M}_{\nu}^{(c)} & =-K_{\nu, c 0} . \quad(c=\alpha, \beta)
\end{align}
By solving the systems of linear Eq.\ref{33} and \ref{34}, we can determine $C_{\gamma i}^{(c)}$ and  $b_{\nu}^{(c)}$. Substituting these into Eq.\ref{32}, we can obtain an approximate S-matrix element
$S_{\gamma c,t}$. Following a procedure similar to that used in the single-channel case, we can ultimately derive a stable S-matrix element $S_{\beta \alpha, s t}$:
\begin{equation}
S_{\beta \alpha, s t}=S_{\beta \alpha, t}-\frac{i k^{2}}{\hbar}\left[\sum_{\gamma} \sum_{i=0}^{n_{\gamma}} K_{\beta 0, \gamma i} C_{\gamma i}^{(\alpha)}+\sum_{\nu} K_{\beta 0, \nu} b_{\nu}^{(\alpha)}\right].
\end{equation}

 After obtaining the scattering matrix elements, how to calculate the scattering phase shifts needs to be discussed under different circumstances. Taking two-channel coupling calculation as an example. (1) Two channels include an open channel and a closed channel. The scattering phase shift of the open channel is affected by the closed channel. After using the aforementioned method to obtain the scattering matrix element of the open channel, substituting it into equation Eq.(\ref{delta}) will yield the scattering phase shift. (2) Both channels are open channels. Four scattering matrix elements ($S_{11}$, $S_{12}$, $S_{21}$, and $S_{22}$) will be obtained, which forms a $2 \times 2$ matrix. Two eigenvalues can be obtained by diagonalizing the $2 \times 2$ matrix, corresponding to the scattering matrix elements of the two open channels. Substituting these two scattering matrix elements into equation Eq.(\ref{delta}) separately will give the scattering phase shifts for both open channels.

\begin{figure}
\centering
\includegraphics[height=10cm,width=15cm]{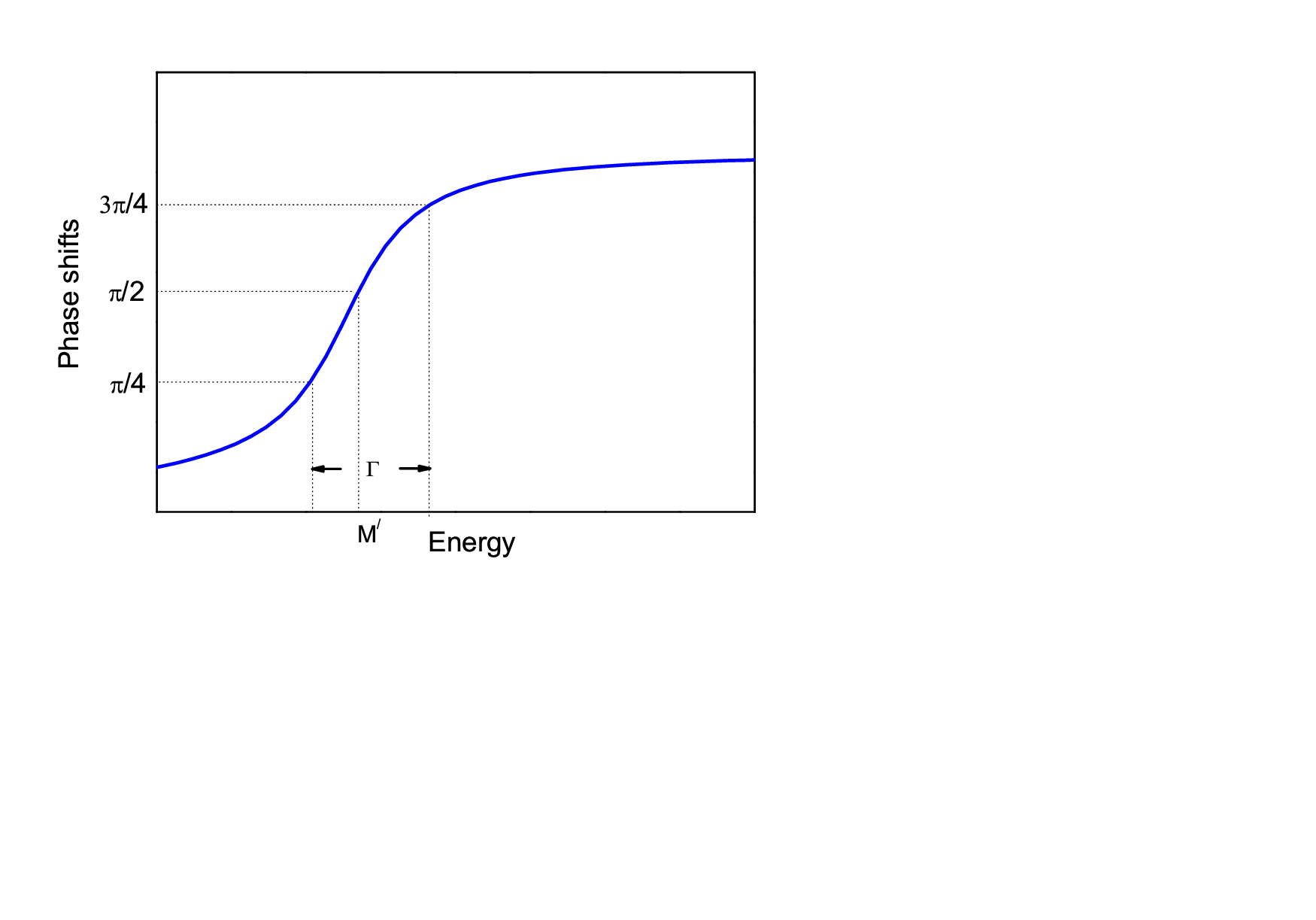}\\
\vspace{-3.5cm}
\caption{The resonance phenomena in scattering phase shifts.} \label{phaseshift}
\end{figure}

Through the scattering process, not only can we better study the interaction between hadrons, but it can also help us research resonance states. The general scattering phase shift diagram should be a smooth curve, that is, the phase shift will change gently as the incident energy increases. But in some cases, the phase shift will be abrupt, the change will be more than 90 degrees, which is the resonance phenomena. The rapid phase change is a general feature of resonance phenomena, see Fig.\ref{phaseshift}. The incident energy at which the phase shift is $\frac{\pi }{2}$ corresponds to the resonance energy ($M^{\prime}$ in Fig.\ref{phaseshift}), and the difference of the energies with phase shift $\frac{3\pi }{4}$ and $\frac{\pi }{4}$ gives the partial decay width of the resonance ($\Gamma$ in Fig.\ref{phaseshift}).


\section{The results and discussions}
In this work, we perform a systematical investigation of the S-wave doubly charmed dibaryon systems with strange $S=0,-2,-4$, isospin $I=0$, and the angular momentum $J=1$. To study the interaction between two hadrons, we calculate the effective potential of the system. Then, a dynamic calculation are carried out to search for bound states. Besides, the scattering process is also investigated to look for the existence of any resonance states.

\subsection{Effective potentials}
\begin{figure}
\begin{center}
\includegraphics[width=3.5in]{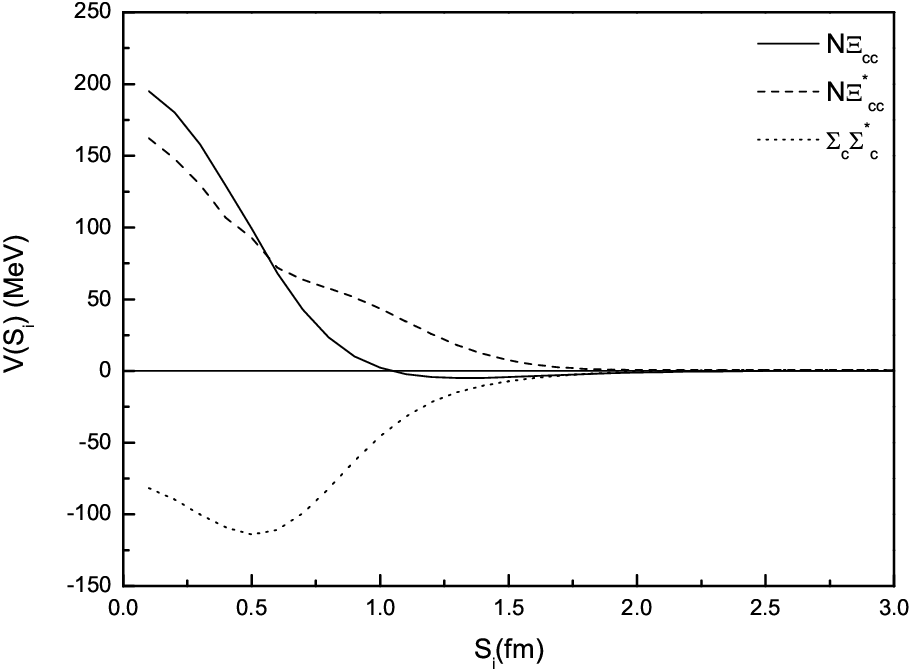}\\
\caption{The effective potentials of different channels of the doubly charmed dibaryon with $S=0$.} \label{C2S0}
\end{center}
\end{figure}

\begin{figure}
\begin{center}
\includegraphics[width=3.5in]{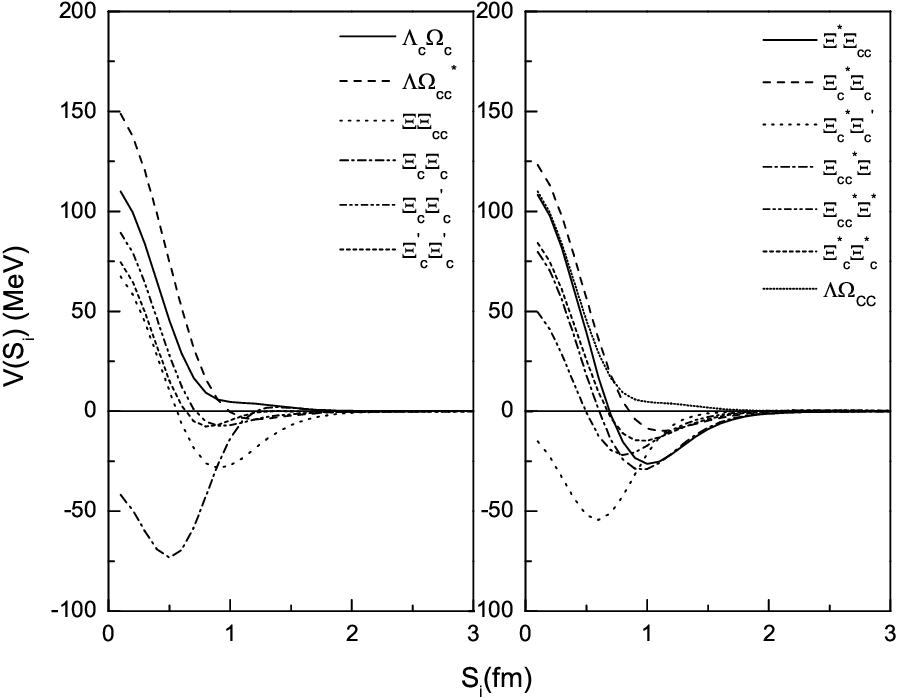}\\
\caption{The effective potentials of different channels of the doubly charmed dibaryon with $S=-2$.} \label{C2S2}
\end{center}
\end{figure}

\begin{figure}
\centering
\vspace{-1cm}
\includegraphics[height=10cm,width=15cm]{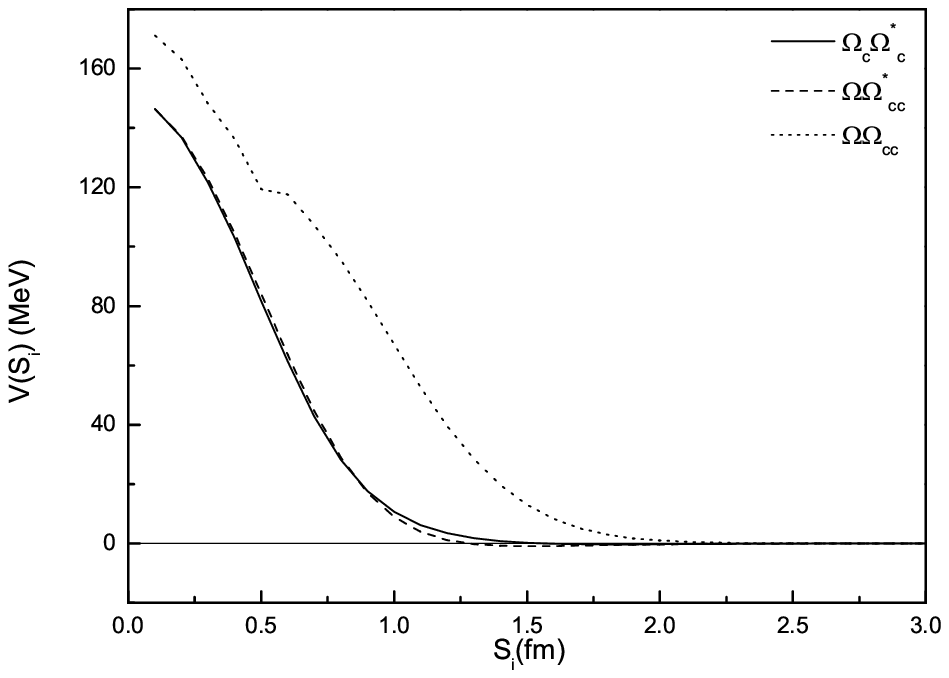} \\
\hspace{-1.0cm}
\vspace{-4.3cm}
\caption{The effective potentials of different channels of the doubly charmed dibaryon with $S=-4$.} \label{C2S4}
\end{figure}

The effective potential between two baryons is shown as
\begin{equation}
V({S_{i}})=E({S_{i}})-E({\infty})
\end{equation}
where $S_{i}$ stands for the distance between two clusters and $E(\infty)$ stands for a sufficient large distance of two clusters, and the expression of $E(S_{i})$ is as follow.
\begin{equation}
E(S_{i})=\frac{\left \langle \Psi _{6q}(S_{i})\left | H \right | \Psi _{6q}(S_{i}) \right \rangle }{\left \langle \Psi _{6q}(S_{i})|\Psi _{6q}(S_{i})  \right \rangle  }
\end{equation}
$\Psi _{6q}(S_{i})$ represents the wave function of a certain channel. Besides, $ \left \langle \Psi _{6q}(S_{i})\left | H \right | \Psi _{6q}(S_{i}) \right \rangle $ and $ \left \langle \Psi _{6q}(S_{i})|\Psi _{6q}(S_{i})  \right \rangle $ are the Hamiltonian matrix and the overlap of the states.
The effective potentials of all channels with different strange numbers are shown in Fig.\ref{C2S0}, Fig.\ref{C2S2} and Fig.\ref{C2S4} respectively.

For $S=0$ system, as shown in Fig.\ref{C2S0}, there are three channels, among which the $N\Xi_{cc}^{\ast}$ channel is repulsive and the other two channels are attractive, one is $N\Xi_{cc}$ channel with weak attraction and the other is $\Sigma_{c}\Sigma_{c}^{*}$ channel with stronger attraction, which indicates that the $\Sigma_{c}\Sigma_{c}^{*}$ channel is more likely to form a bound state or resonance state.

For $S=-2$ system, it is a complicated system with 13 channels. From Fig.\ref{C2S2} we can see that the potentials of the $\Lambda_{c}\Omega_{c}$ and $\Lambda\Omega_{cc}$ are repulsive, while the potentials for the other 11 channels are attractive. Among these 11 channels, the attraction of $\Xi_{c}\Xi_{c}$ and $\Xi_{c}^{*}\Xi_{c}^{\prime}$ is much stronger than that of other attractive channels, which implies that it is more possible for $\Xi_{c}\Xi_{c}$ and $\Xi_{c}^{*}\Xi_{c}^{\prime}$ to form bound states or resonance states.

For $S=-4$ system, seen in Fig.\ref{C2S4}, there are only three channels, which are $\Omega_{c}\Omega_{c}^{\ast}$, $\Omega\Omega_{cc}^{\ast}$ and $\Omega\Omega_{cc}$. The potentials of these channels all  purely repulsive. Therefore, it is difficult for these channels to form any bound state. However, we still need to confirm the existence of bound states or resonance states by performing the dynamic calculations.

\begin{figure}[ht]
\begin{flushleft}
\includegraphics[width=3.5in]{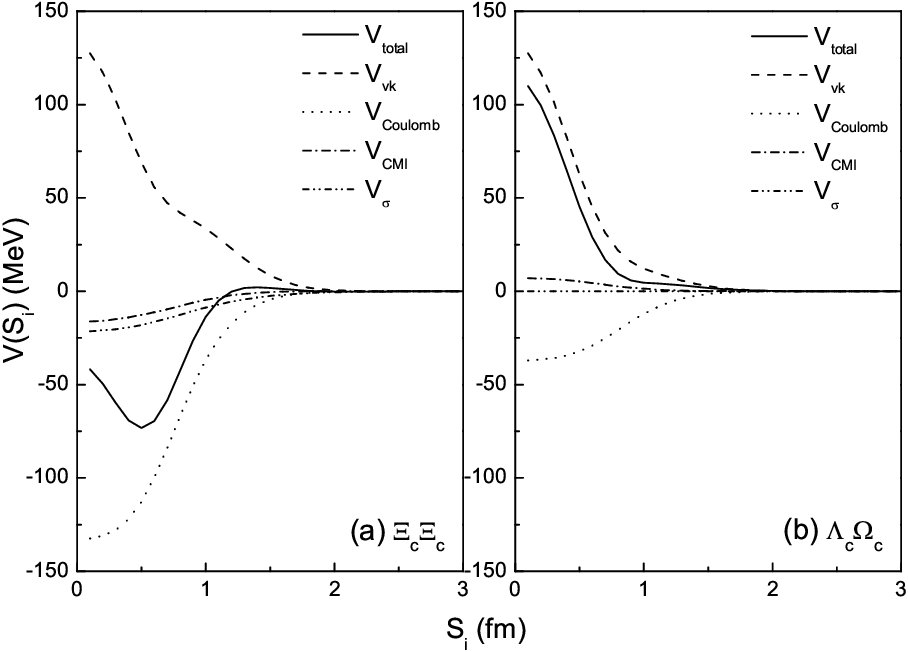}\\
\caption{The contributions to the effective potential from various interaction terms of the doubly charmed dibaryon with $S=-2$.} \label{every}
\end{flushleft}
\end{figure}

In addition, to investigate the interactions between two baryons in detail, we calculate the contribution of each interaction term to the effective potential, including the kinetic energy ($V_{vk}$), the confinement ($V_{con}$), the Coulomb ($V_{Coulomb}$), the chromomagnetic interaction ($V_{CMI}$), the one-boson exchange ($V_{\pi}$, $V_{K}$ and $V_{\eta}$), the heavy meson exchange ($V_{D}$, $V_{D_{s}}$ and $V_{\eta_{c}}$) and the $\sigma$-meson exchange ($V_{\sigma}$).
To save space, we take two doubly charmed dibaryon channel with $S=-2$ system as an example here. One is the channel $\Xi_{c}\Xi_{c}$ with a deep attractive potential, and the other is the channel $\Lambda_{c}\Omega_{c}$ with a pure repulsive potential, as shown in Fig.~\ref{every}.
Due to the fact that both baryon clusters are color singlet, the confinement potential between them is zero.
Additionally, since the contributions of the meson exchange ($\pi, K, \eta, D, D_{s}$ and $D_{s}$) interactions to the effective potentials are relatively small, they are not shown in Fig.~\ref{every}.
For the $\Xi_{c}\Xi_{c}$ channel, shown in Fig.~\ref{every}(a), although the kinetic energy term provides
repulsive interaction, the attraction provided by Coulomb potential is quite significant, while the potentials of the CMI and the $\sigma$-meson exchange also provide weak attraction, resulting in a deep attraction in the overall interaction of the $\Xi_{c}\Xi_{c}$ channel. For the $\Lambda_{c}\Omega_{c}$ channel, shown in Fig.~\ref{every}(b), although the Coulomb potential provides some attraction, it is not sufficient, being less than 50 MeV. The kinetic energy term and the CMI provides repulsive interaction.
In addition, since the $\sigma$-meson only exchanges between u and d quarks, it does not work between $\Lambda_{c}$ and $\Omega_{c}$. Thus, this channel exhibits a repulsive effective potential.

\subsection{Bound state calculation}
In order to see whether there is any bound state, a dynamic calculation based on RGM~\cite{RGM1} has been performed. The energies of each channel as well as the one with channel coupling calculation are listed in Table~\ref{result1}, Table~\ref{result2} and Table~\ref{result3}. The first column is the state of every channel; the second column $E_{th}$ denotes the theoretical threshold of each corresponding state; the third column $E_{sc}$ represents the energy of every single channel; the fourth column $B_{sc}$ stands for the binding energy of every single channel, which is $B_{sc}=E_{sc}-E_{th}$; the fifth column $E_{cc}$ denotes the lowest energy of the system by channel coupling calculation; and the last column $B_{cc}$ represents the binding energy with all channels coupling, which is $B_{cc}= E_{cc}-E_{th}$. Here, we should notice that when the state is unbound, we label it as ``ub". \\

\begin{table*}[ht]
\renewcommand{\arraystretch}{1.5}
\caption{The energy (in MeV) of S=0 for the charmed dibaryon systems.}   \label{result1}
\centering
\begin{tabular}{ccccccc}
\hline \hline
   ~~ $Channels$ &  ~~$E_{th}$~(MeV) &  ~~$E_{sc}$~(MeV) &  ~~$B_{sc}$~(MeV) &    ~~$E_{cc}$~(MeV) &  ~~$B_{cc}$~(MeV)
  \\ \hline
    ~~$N\Xi_{cc}$ &    ~~4425.83 &    ~~4431.70&   ~~ub &   ~~\multirow{3}{*}{4431.61} &  ~~\multirow{3}{*}{ub}    \\ \cline{1-4}
    ~~$N\Xi^{\ast}_{cc}$ &      ~~4458.63 &   ~~4466.05&    ~~ub &     ~~ &      \\ \cline{1-4}
   ~~$\Sigma_{c}\Sigma^{\ast}_{c}$ &   ~~4865.68 &   ~~4820.35 &    ~~-45.33  &     ~~ &      \\ \hline\hline
\end{tabular}
\end{table*}

\begin{table*}[ht]
\renewcommand{\arraystretch}{1.5}
\caption{The energy (in MeV) of S=-2 for the charmed dibaryon systems.}   \label{result2}
\centering
\begin{tabular}{ccccccc}
\hline \hline
   ~~ $Channels$ &  ~~$E_{th}$~(MeV) &  ~~$E_{sc}$~(MeV) &  ~~$B_{sc}$~(MeV) &    ~~$E_{cc}$~(MeV) &  ~~$B_{cc}$~(MeV)
  \\ \hline
 ~~${\Lambda_{c}\Omega_{c}}$ &    ~~4920.59&     ~~4925.48 &   ~~ub &   ~~\multirow{13}{*}{4741.10} &   ~~\multirow{13}{*}{ub}   \\ \cline{1-4}
    ~~$\Lambda\Omega^{\ast}_{cc}$ &   ~~4753.46 &   ~~4758.12 &    ~~ub &    ~~ &     \\ \cline{1-4}
    ~~$\Xi\Xi_{cc}$ &   ~~4830.97 &   ~~4831.83 &    ~~ub &     ~~ &     \\ \cline{1-4}
~~$\Xi_{c}\Xi_{c}$ &   ~~4900.84 &   ~~4879.32 &    ~~-21.52  &     ~~ &     \\\cline{1-4}
~~$\Xi_{c}\Xi^{\prime}_{c}$ &   ~~4996.49 &   ~~5000.57 &    ~~ub  &     ~~ &     \\\cline{1-4}
~~$\Xi^{\prime}_{c}\Xi^{\prime}_{c}$ &   ~~5092.14 &   ~~5096.01 &    ~~ub  &     ~~ &     \\\cline{1-4}
~~$\Xi^{*}\Xi_{cc}$ &   ~~4999.74 &   ~~5001.45 &    ~~ub  &     ~~ &     \\\cline{1-4}
~~$\Xi_{c}\Xi^{\ast}_{c}$ &   ~~5021.79 &   ~~5025.93 &    ~~ub  &     ~~ &     \\\cline{1-4}
~~$\Xi^{*}_{c}\Xi^{\prime}_{c}$ &   ~~5117.44&   ~~5107.43 &    ~~-10.01 &    ~~ &     \\\cline{1-4}
~~$\Xi^{*}_{cc}\Xi$ &   ~~4863.77&   ~~4864.49 &    ~~ub &    ~~ &     \\\cline{1-4}
~~$\Xi^{*}_{cc}\Xi^{*}$ &   ~~5032.54&   ~~5035.12 &    ~~ub &    ~~ &     \\\cline{1-4}
~~$\Xi^{*}_{c}\Xi^{*}_{c}$ &   ~~5142.74&   ~~5146.35 &    ~~ub &    ~~ &     \\\cline{1-4}
~~$\Lambda\Omega_{cc}$ &   ~~4736.79&   ~~4741.03 &    ~~ub &    ~~ &     \\
 \hline\hline
\end{tabular}
\end{table*}

\begin{table*}[ht]
\renewcommand{\arraystretch}{1.5}
\caption{The energy (in MeV) of S=-4 for the charmed dibaryon systems.}   \label{result3}
\centering
\begin{tabular}{ccccccc}
\hline \hline
   ~~ $Channels$ &  ~~$E_{th}$~(MeV) &  ~~$E_{sc}$~(MeV) &  ~~$B_{sc}$~(MeV) &    ~~$E_{cc}$~(MeV) &  ~~$B_{cc}$~(MeV)
  \\ \hline
    ~~$\Omega_{c}\Omega^{\ast}_{c}$ &    ~~5409.71 &    ~~5413.42&   ~~ub &   ~~\multirow{3}{*}{5303.93} &  ~~\multirow{3}{*}{ub}    \\ \cline{1-4}
    ~~$\Omega\Omega^{\ast}_{cc}$ &      ~~5317.31 &   ~~5320.97&    ~~ub &     ~~ &      \\ \cline{1-4}
   ~~$\Omega\Omega_{cc}$ &   ~~5299.52 &   ~~5303.98 &    ~~ub  &     ~~ &      \\ \hline\hline
\end{tabular}
\end{table*}

 \textbf{\textit{S=0}}:
  The single channel calculation shows that the channel $\Sigma_{c}\Sigma^{\ast}_{c}$ is bound states with the binding energy -45~MeV (see Table~\ref{result1}). This conclusion is consistent with the property that there is a strong effective attraction of this channel.
  However, for the $N\Xi_{cc}$ and $N\Xi^{\ast}_{cc}$ channels, the energies are above their corresponding thresholds due to the weak attraction and repulsive of these channels, so neither of these channels are bound. For the calculation of the channel coupling, the lowest energy is still above the lowest threshold ($N\Xi_{cc}$). Therefore, for this system, no any bound state below the lowest threshold is found. For the higher-energy single-channel bound state $\Sigma_{c}\Sigma^{\ast}_{c}$, it can couple with the open channels and the scattering process is needed to determine whether the $\Sigma_{c}\Sigma^{\ast}_{c}$ is a resonance state or not, which will be discussed in subsection C.

 \textbf{\textit{S=-2}}:
 From Table~\ref{result2}, it shows that both $\Xi_{c}\Xi_{c}$ and $\Xi^{*}_{c}\Xi^{\prime}_{c}$ channels are bound with the bounding energies -22~MeV and -10~MeV, respectively, while the other 11 channels are unbound. This is in good agreement with the result of the effective potentials.
After the channel coupling calculation, the lowest energy of this system is 4741 MeV (still higher than the threshold of the lowest channel $\Lambda\Omega_{cc}$ ), which indicates that there is no any bound state below the lowest threshold for this system. Further calculation of scattering process is needed to find any resonance state.

\textbf{\textit{S=-4}}: As shown in Table~\ref{result3}, all of the channels $\Omega_{c}\Omega^{\ast}_{c}$, $\Omega\Omega^{\ast}_{cc}$ and $\Omega\Omega_{cc}$ are unbound. This is reasonable. The attractions of all these three channels are purely repulsive as shown in Fig.\ref{C2S4}, so they can not form any bound state. The lowest energy of the system is still higher than the lowest threshold of the $\Omega\Omega_{cc}$ channel by the channel coupling calculation, which indicates that the system with $S=-4$ is unbound.

\subsection{Resonance states}
As mentioned above, some channels are bound due to the strong attractions of the system. However, these states will decay to the corresponding open channels by coupling with them and become resonance states. Besides, some states will become scattering state by the effect of coupling to both the open and closed channels. To further check the existence of the resonance states, we studied the scattering phase shifts of all possible open channels. Since no resonance states are obtained in the $S=-4$ system, we only show the scattering phase shifts of the $S=0$ and $S=-2$ systems here.

In the $S=0$ system, only one bound states $\Sigma_{c}\Sigma^{\ast}_{c}$ is obtained and there are two open channels $N\Xi_{cc}$ and $N\Xi^{\ast}_{cc}$. We analyze the two-channel coupling with the bound state and a related open channel. The general features of the calculated results are as follows.

\begin{figure}
\begin{flushleft}
\includegraphics[width=3.5in]{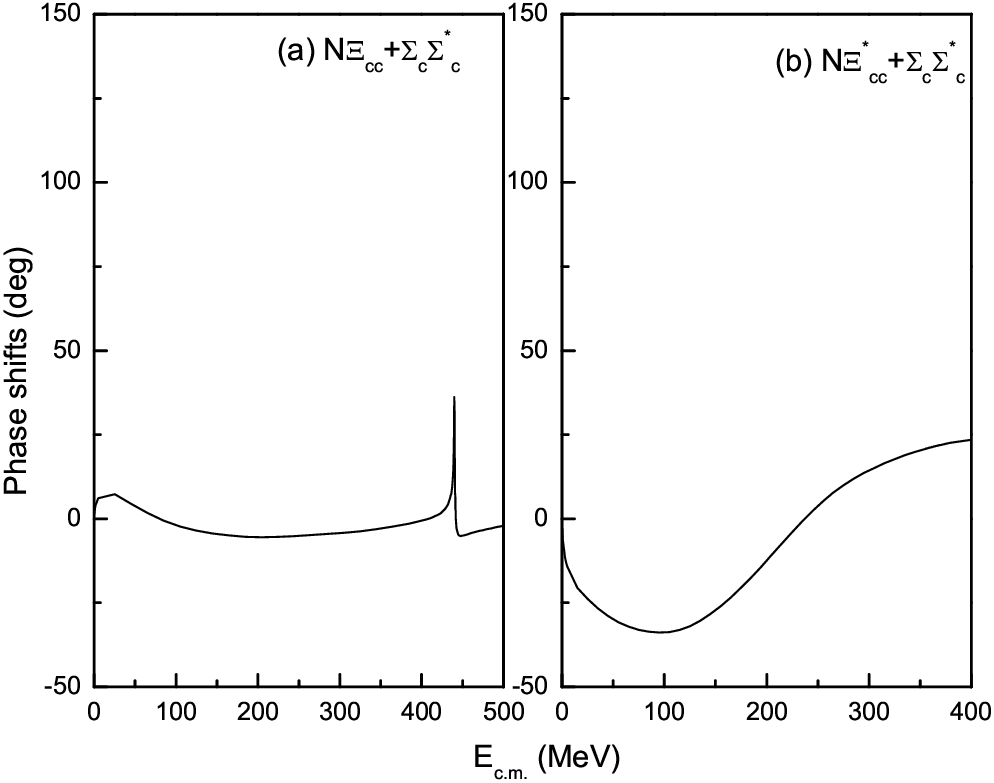}\\
\caption{The phase shift with two-channel coupling for the $S=0$ system.} \label{C2S0-phase}
\end{flushleft}
\end{figure}

In $N\Xi_{cc}$ scattering process, as shown in Fig.\ref{C2S0-phase}(a), $\Sigma_{c}\Sigma^{\ast}_{c}$ does not behave as resonance state. However, at the point where the incident energy is around 440 MeV, a cusp appears, and with the addition of the threshold for channel $N\Xi_{cc}$, the energy corresponds exactly to the threshold for channel $\Sigma_{c}\Sigma^{\ast}_{c}$. This indicates that at the cusp point, channel $\Sigma_{c}\Sigma^{\ast}_{c}$ has become an open channel.
The reason why no resonance state appears is that $\Sigma_{c}\Sigma^{\ast}_{c}$ channel is pushed above the threshold value with the influence of the channel coupling effect. At the same time, as shown in in Fig.\ref{C2S0-phase}(b), no sharp increase in phase shift as the incident energy increases. We examined a series of the cross matrix elements between $N\Xi^{\ast}_{cc}$ and $\Sigma_{c}\Sigma^{\ast}_{c}$ they are all close to zero, which means that the coupling between $N\Xi^{\ast}_{cc}$ and $\Sigma_{c}\Sigma^{\ast}_{c}$ is very weak, so it is difficult to form resonance state.

\begin{figure}
\begin{flushleft}
\includegraphics[width=3.5in]{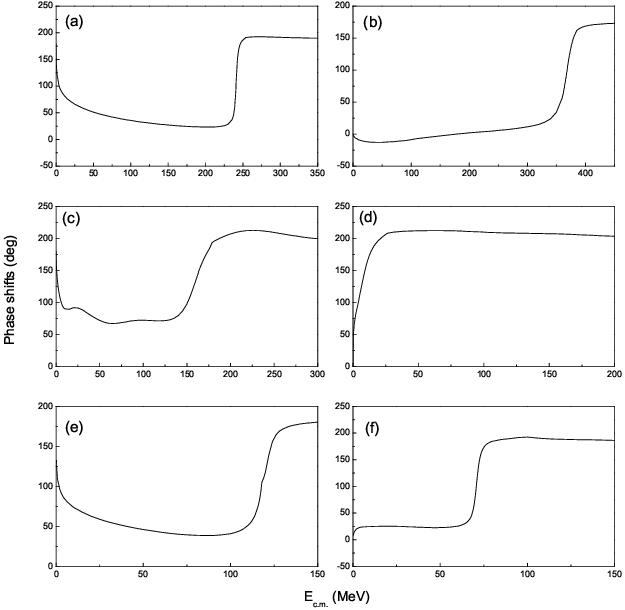}\\
\caption{The phase shift with two-channel coupling for the $S=-2$ system. (a)$\Xi_{cc}^*\Xi$+$\Xi_{c}^{\ast}\Xi_{c}^{\prime}$, (b)$\Lambda\Omega_{cc}$+$\Xi_{c}^{\ast}\Xi_{c}^{\prime}$, (c)$\Lambda_{c}\Omega_{c}$+$\Xi_{c}^{\ast}\Xi_{c}^{\prime}$, (d)$\Xi_{c}^{\prime}\Xi_{c}^{\prime}$+$\Xi_{c}^{\ast}\Xi_{c}^{\prime}$,
(e)$\Xi^{\ast}\Xi_{cc}$+$\Xi_{c}^{\ast}\Xi_{c}^{\prime}$, (f)$\Xi_{c}\Xi_{c}^{\ast}$+$\Xi_{c}^{\ast}\Xi_{c}^{\prime}$} \label{C2S2-2}
\end{flushleft}
\end{figure}

Here, we should note that the horizontal axis $E_{c.m.}$ in Fig.\ref{C2S0-phase} is the incident energy without the theoretical threshold of the corresponding open channel. So the theoretical resonance mass $M^{\prime}$ is obtained by adding $E_{c.m.}$ and the theoretical threshold of the corresponding open channel. In order to minimize the theoretical errors and compare our predictions with future experimental data, we shift the theoretical resonance mass by $M=M^{\prime}-E_{th}+E_{exp}$, where $E_{th}$ and $E_{exp}$ are the theoretical and experimental thresholds of the resonance state, respectively. Taking the resonance state $\Xi^{*}_{c}\Xi^{\prime}_{c}$ in the $\Xi^{*}_{cc}\Xi$ channel as an example, the resonance mass shown in Fig.\ref{C2S2-2}(a) is $M^{\prime}=5104$ MeV, the theoretical threshold is $M_{th}=5117$ MeV, and the experimental threshold is $M_{exp}=5223$ MeV. Then the final experimental resonance mass $M=5104-5117+5223=5210$ MeV. In the following discussion, we will directly provide the resonance mass after the mass shift.

In the $S=-2$ system, as is listed in Table~\ref{result2}, there are two bound states in the single channel calculation, which are $\Xi_{c}\Xi_{c}$ and $\Xi^{*}_{c}\Xi^{\prime}_{c}$. It is worth mentioning that we analyze two types of channel coupling in this work.
The first is the two-channel coupling with a singly bound state and a related open channel, while the
other is the multi-channel coupling with two bound states and the corresponding open channels. The general features of the calculated results are as follows. Before calculating the scattering phase shifts, the cross matrix elements were first computed. It was found that for the closed channel $\Xi_{c}\Xi_{c}$, the cross matrix elements with the channels $\Lambda\Omega_{cc}^{\ast}$ and $\Xi\Xi_{cc}$ are 0. Therefore, for the closed channel $\Xi_{c}\Xi_{c}$, its open channels are $\Xi_{cc}^{\ast}\Xi$ and $\Lambda\Omega_{cc}$. For the closed channel $\Xi^{\ast}_{c}\Xi^{\prime}_{c}$, the cross matrix elements with the channels $\Lambda\Omega_{cc}^{\ast}$, $\Xi\Xi_{cc}$, $\Xi_{c}\Xi^{\prime}_{c}$, and $\Xi^{\ast}_{cc}\Xi^{\ast}$ are also zero. Hence, the other six channels $\Lambda\Omega_{cc}$, $\Xi_{cc}^{\ast}\Xi$, $\Lambda_{c}\Omega_{c}$, $\Xi^{\prime}_{c}\Xi^{\prime}_{c}$, $\Xi^{\ast}\Xi_{cc}$, and $\Xi_{c}\Xi^{\ast}_{c}$ will serve as the open channels for the closed channel $\Xi^{*}_{c}\Xi^{\prime}_{c}$.

\begin{figure}
\begin{flushleft}
\includegraphics[width=3.5in]{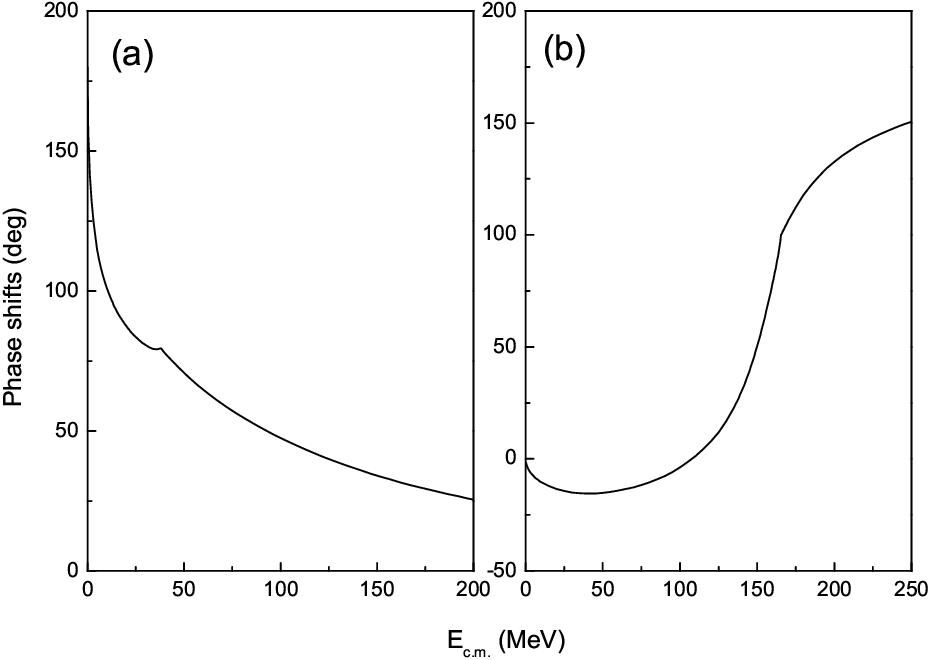}\\
\caption{The phase shift with two-channel coupling for the $S=-2$ system. (a)$\Xi_{cc}^*\Xi$+$\Xi_{c}\Xi_{c}$, (b)$\Lambda\Omega_{cc}$+$\Xi_{c}\Xi_{c}$}\label{C2S2-1}
\end{flushleft}
\end{figure}

To search for the $\Xi_{c}\Xi_{c}$ resonance state, the scattering phase shifts of the open channels $\Xi_{cc}^{\ast}\Xi$ and $\Lambda\Omega_{cc}$ are calculated respectively, as shown in Fig.\ref{C2S2-1}.
From Fig.\ref{C2S2-1}(a), one can see that the scattering phase shifts of the $\Xi_{cc}^{\ast}\Xi$ channel changes smoothly as the incident energy increases, indicating that the $\Xi_{c}\Xi_{c}$ becomes a scattering state after coupling with $\Xi_{cc}^{\ast}\Xi$, although $\Xi_{c}\Xi_{c}$ is a bound state in the single channel calculation. For the open channel $\Lambda\Omega_{cc}$, as shown in Fig.\ref{C2S2-1}(b), the phase shifts exhibit a resonance state with a decay width 57.0 MeV. This indicates that $\Xi_{c}\Xi_{c}$ behaves as a resonance state in the $\Lambda\Omega_{cc}$ scattering process. The theoretical resonance mass is 4898 MeV, and the experimental resonance mass is 4938 MeV after the mass shift. Although the $\Xi_{c}\Xi_{c}$ exhibits a resonance state in the scattering process with the open channel $\Lambda\Omega_{cc}$, the $\Xi_{c}\Xi_{c}$ disappears in the $\Xi_{cc}^{\ast}\Xi$ scattering phase shifts, it will decay through the $\Xi_{cc}^{\ast}\Xi$ open channel. So the $\Xi_{c}\Xi_{c}$ cannot be identified as a resonance state.

To search for the $\Xi_{c}^{\ast}\Xi_{c}^{\prime}$ resonance state, the scattering phase shifts of six open channels $\Xi_{cc}^{\ast}\Xi$, $\Lambda\Omega_{cc}$, $\Lambda_{c}\Omega_{c}$, $\Xi_{c}^{\prime}\Xi_{c}^{\prime}$, $\Xi^{\ast}\Xi_{cc}$ and $\Xi_{c}\Xi_{c}^{\ast}$ are calculated respectively, as shown in Fig.\ref{C2S2-2}.
It is obvious that the $\Xi_{c}^{\ast}\Xi_{c}^{\prime}$ performs as a resonance state in all these six open channels. The resonance masses and decay widths of the $\Xi_{c}^{\ast}\Xi_{c}^{\prime}$ in above six open channels are: $M_1=5212$ MeV, $\Gamma_{1}=5.7$ MeV; $M_2=5212$ MeV, $\Gamma_{2}=23.0$ MeV; $M_3=5187$ MeV, $\Gamma_{3}=34.5$ MeV; $M_4=5203$ MeV, $\Gamma_{4}=9.6$ MeV; $M_5=5224$ MeV, $\Gamma_{5}=7.7$ MeV; $M_6=5199$ MeV, $\Gamma_{6}=4.7$ MeV; respectively. Finally, the total decay width of this resonance state is 84.6 MeV. However, further calculations of multi-channel coupling are needed to confirm whether this state is a resonance state or not.

For the case of multi-channel coupling, considering two closed channels $\Xi_{c}\Xi_{c}$ and $\Xi_{c}^{\ast}\Xi_{c}^{\prime}$, their common open channels are $\Lambda\Omega_{cc}$ and $\Xi_{cc}^{\ast}\Xi$. To further investigate the effect of channel coupling, we will calculate the following cases separately.

\begin{figure}
\begin{flushleft}
\includegraphics[width=3.0in]{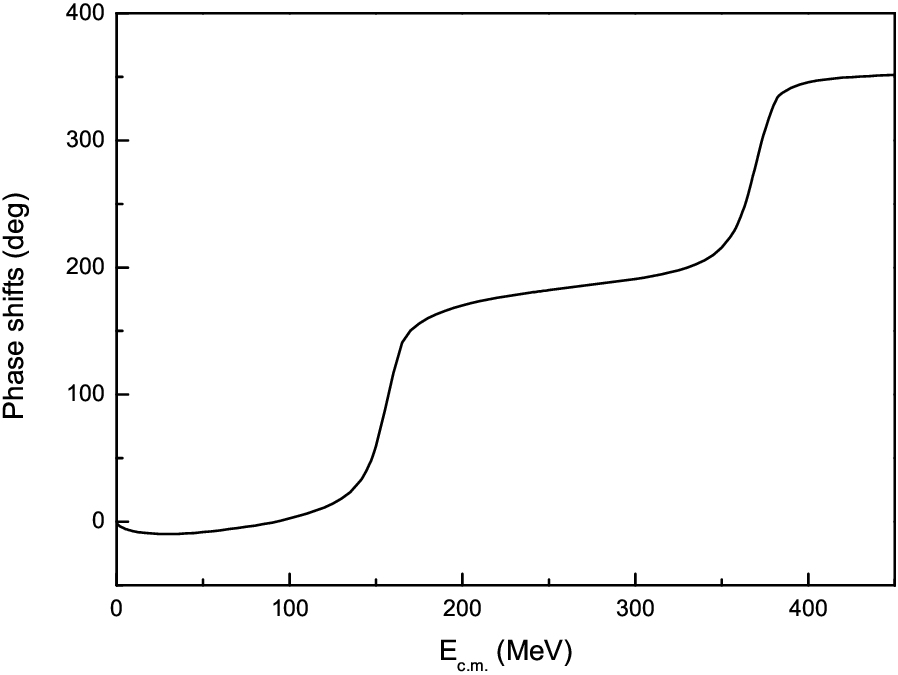}\\
\caption{The phase shift with multi-channel coupling ($\Lambda\Omega_{cc}$+$\Xi_{c}\Xi_{c}$+$\Xi_{c}^{\ast}\Xi_{c}^{\prime}$) for the $S=-2$ system.} \label{C2S2-CC1}
\end{flushleft}
\end{figure}

$\bullet$ $\Lambda\Omega_{cc}$+$\Xi_{c}\Xi_{c}$+$\Xi_{c}^{\ast}\Xi_{c}^{\prime}$: The phase shifts of the open channel $\Lambda\Omega_{cc}$ coupling with two closed channels $\Xi_{c}\Xi_{c}$ and $\Xi_{c}^{\ast}\Xi_{c}^{\prime}$ are calculated, which are shown in Fig.\ref{C2S2-CC1}. It is obvious that as the incident energy increases, there are two abrupt changes in the scattering phase shift, indicating the possible existence of two resonance states during this scattering process, which are $\Xi_{c}\Xi_{c}$ and $\Xi_{c}^{\ast}\Xi_{c}^{\prime}$. The resonance masses and decay widths are 4932 MeV and 19.6 MeV for $\Xi_{c}\Xi_{c}$; 5213 MeV and 18.8 MeV for $\Xi_{c}^{\ast}\Xi_{c}^{\prime}$, respectively.

\begin{figure}
\begin{flushleft}
\includegraphics[width=3.0in]{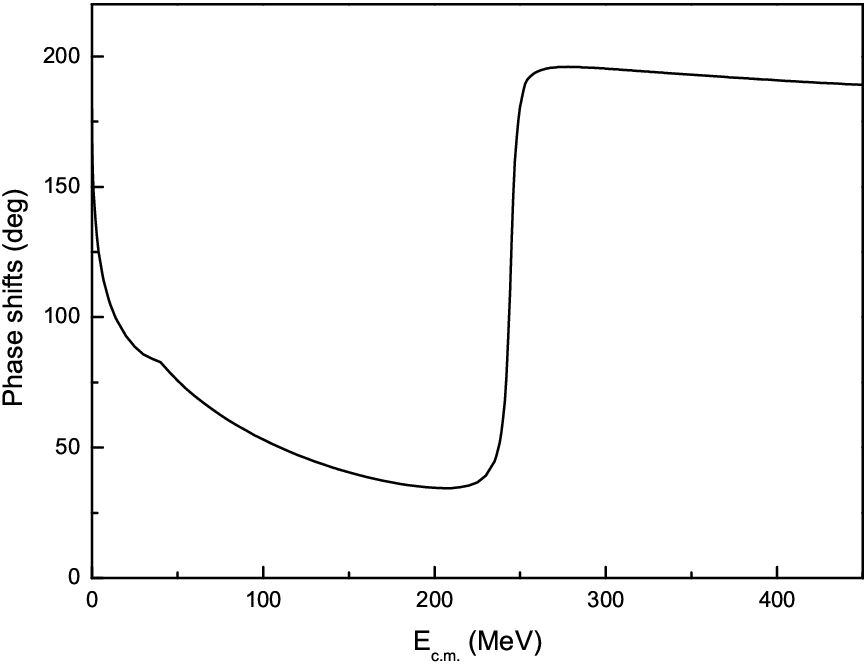}\\
\caption{The phase shift with multi-channel coupling ($\Xi_{cc}^{\ast}\Xi$+$\Xi_{c}\Xi_{c}$+$\Xi_{c}^{\ast}\Xi_{c}^{\prime}$) for the $S=-2$ system.} \label{C2S2-CC2}
\end{flushleft}
\end{figure}

$\bullet$ $\Xi_{cc}^{\ast}\Xi$+$\Xi_{c}\Xi_{c}$+$\Xi_{c}^{\ast}\Xi_{c}^{\prime}$: The phase shifts of the open channel $\Xi_{cc}^{\ast}\Xi$ coupling with two closed channels $\Xi_{c}\Xi_{c}$ and $\Xi_{c}^{\ast}\Xi_{c}^{\prime}$ are calculated, which are shown in Fig.\ref{C2S2-CC2}. It can be seen that the phase shifts of the $\Xi_{cc}^{\ast}\Xi$ are different from those of the $\Lambda\Omega_{cc}$ channels. When the incident energy is close to 0, the scattering phase shift approaches to 180 degree, indicating that the $\Xi_{cc}^{\ast}\Xi$ channel becomes a bound state with the effect of the channel coupling. As the incident energy increases, a resonance state $\Xi_{c}^{\ast}\Xi_{c}^{\prime}$ is obtained with the resonance mass and decay width 5214 MeV and 5.82 MeV. However, the $\Xi_{c}\Xi_{c}$ state transforms into a scattering state with the effect of the channel coupling.

$\bullet$ $\Lambda\Omega_{cc}$+$\Xi_{cc}^{\ast}\Xi$+$\Xi_{c}\Xi_{c}$+$\Xi_{c}^{\ast}\Xi_{c}^{\prime}$: Since $\Xi_{cc}^{\ast}\Xi$ becomes a bound state after the above three channel coupling, we study the scattering process of the open channel $\Lambda\Omega_{cc}$ coupling with three closed channels $\Xi_{cc}^{\ast}\Xi$, $\Xi_{c}\Xi_{c}$, and $\Xi_{c}^{\ast}\Xi_{c}^{\prime}$. There are two sharp changes approximately 180 degree in the $\Lambda\Omega_{cc}$ channel scattering process.
According to the above analysis, after the channel coupling, $\Xi_{cc}^{\ast}\Xi$ becomes a bound state while $\Xi_{c}\Xi_{c}$ transforms into a scattering state. Besides, the theoretical resonance mass for the first resonance is $M^{\prime}=4858$ MeV, lower the threshold of $\Xi_{cc}^{\ast}\Xi$. Therefore, the first resonance is predominantly constituted by $\Xi_{cc}^{\ast}\Xi$.
The experimental resonance mass is 5081 MeV and the decay width is 0.3 MeV. It should be noted that due to the absence of experimental observations of the $\Xi_{cc}^{\ast}$ state, the experimental value for $\Xi_{cc}^{\ast}$ utilized in this context is derived from lattice QCD~\cite{Lewis:2001iz}. The theoretical resonance mass for the second resonance is $M^{\prime}=5107$ MeV, slightly below the threshold of $\Xi_{c}^{\ast}\Xi_{c}^{\prime}$, hence the second resonance state is $\Xi_{c}^{\ast}\Xi_{c}^{\prime}$. The final experimental resonance mass and decay width are 5213 MeV and 19.8 MeV, respectively.

\begin{figure}
\centering
\includegraphics[width=3.0in]{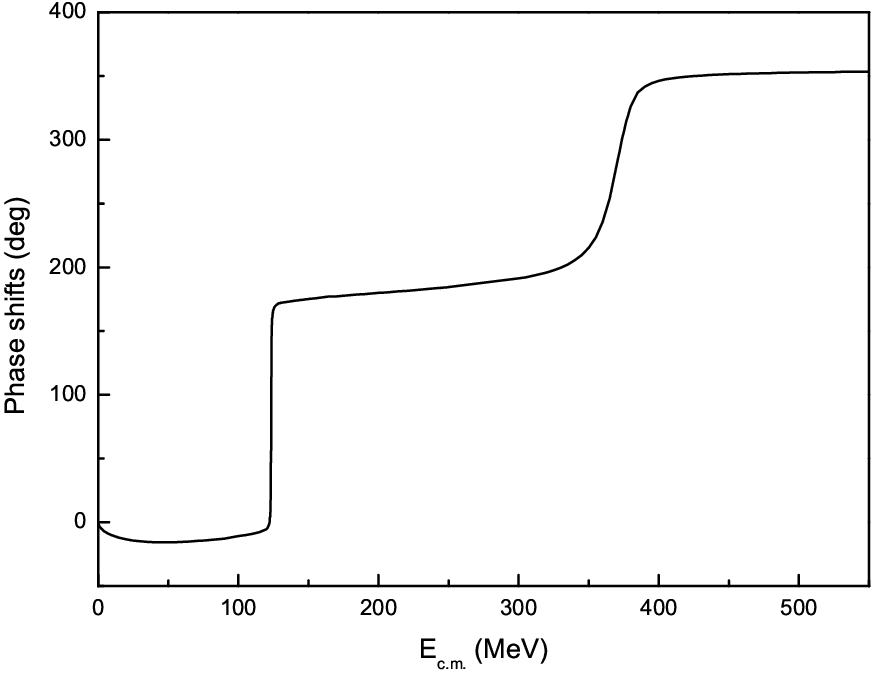}\\
\caption{The phase shift with multi-channel coupling ($\Lambda\Omega_{cc}$+$\Xi_{cc}^{\ast}\Xi$+$\Xi_{c}\Xi_{c}$+$\Xi_{c}^{\ast}\Xi_{c}^{\prime}$)for the $S=-2$ system.} \label{C2S2-CC3}
\end{figure}

\section{Summary}
The S-wave doubly charmed dibaryon systems with strangeness numbers $S=0$, $-2$ and $-4$ are systemically investigated by using the RGM in the framework of ChQM. Our goal is to search for any bound state or resonance state of doubly charmed dibaryon systems. Herein, the effective potentials are calculated to explore the interactions between two baryons. Both the single-channel and the coupled-channel dynamic bound-state calculations are carried out to search for possible states. Meanwhile, the study of the scattering process of the open channels is carried out to confirm possible resonance states.

For the $S=0$ system, the attractions between $\Sigma_{c}\Sigma_{c}^{*}$ are large enough to form singly bound states. However, after coupling with the corresponding open channels $N\Xi_{cc}$ and $N\Xi_{cc}^{\prime}$, $\Sigma_{c}\Sigma_{c}^{*}$ becomes a scattering state.

For the $S=-2$ system, according to the numerical results, the attractions between $\Xi_{c}$ and $\Xi_{c}$, $\Xi_{c}^{\prime}$ and $\Xi_{c}^{\ast}$ are large enough to form singly bound states $\Xi_{c}\Xi_{c}$ and  $\Xi_{c}\Xi_{c}^{\ast}$. However, these states can couple with the corresponding open channels, and become resonance states or scattering states. By including the effect of channel-coupling, two resonance states are obtained, which are $\Xi_{cc}^{\ast}\Xi$ with resonance mass and decay width 5081 MeV and 0.3 MeV, and the $\Xi_{c}\Xi_{c}^{\ast}$ state with the mass and width 5213 MeV and 19.8 MeV, respectively. Both states predominantly decay through the $\Lambda\Omega_{cc}$ channel. All these dibaryons are worth searching for in experiments, although it will be a challenging subject.


In this work, fewer charmed dibaryon resonance states are obtained by comparing with the work of various approaches, such as the chiral effective field theory~\cite{Wang:2024riu},  the chromomagnetic interaction model~\cite{Liu:2022rzu}, the QCD Sum Rules~\cite{Wang:2021pua} and so on. More bound states are obtained in the single-channel calculation. However, the observable of these states depends on its coupling to the open channels. The coupling can shift the energy of the resonance, give the width to the resonance and even destroy the resonance. Therefore, to provide the necessary information for experiments to search for exotic hadron states, mass spectrum calculations alone are not enough. The coupling calculation between the bound channels and open channels
is indispensable.

\acknowledgments{This work is supported partly by the National Natural Science Foundation of China under Contracts Nos. 11675080, 11775118 and 11535005.}

\section{Appendix}
 Here, we only list the wave functions we used in this work. The spin wave function of a $q^{3}$ cluster is labeled as $\chi_{s,s_{z}}^{\sigma}$, where $s$ and $s_{z}$ are the spin quantum number and the third component, respectively. For wave functions with the same quantum number but different symmetries, we distinguish them with different numbers. For example, $\chi_{\frac{1}{2},\frac{1}{2}}^{\sigma1}$ and $\chi_{\frac{1}{2},\frac{1}{2}}^{\sigma2}$ represent respectively the symmetric and antisymmetric spin wave functions with spin quantum number $\frac{1}{2}$.
 \begin{eqnarray}
	\chi_{\frac{3}{2}, \frac{3}{2}}^{\sigma} & = &\alpha \alpha \alpha \nonumber \\
	\chi_{\frac{3}{2}, \frac{1}{2}}^{\sigma} & = &\frac{1}{\sqrt{3}}(\alpha \alpha \beta+\alpha \beta \alpha+\beta \alpha \alpha) \nonumber \\
	\chi_{\frac{3}{2},-\frac{1}{2}}^{\sigma} & = &\frac{1}{\sqrt{3}}(\alpha \beta \beta+\beta \alpha \beta+\beta \beta \alpha) \nonumber \\
    \chi_{\frac{3}{2},-\frac{3}{2}}^{\sigma} & = &\beta \beta \beta \nonumber \\
    \chi_{\frac{1}{2},\frac{1}{2}}^{\sigma1} & = &\sqrt{\frac{1}{6}}(2 \alpha \alpha \beta-\alpha \beta \alpha-\beta \alpha \alpha)  \nonumber\\
    \chi_{\frac{1}{2},\frac{1}{2}}^{\sigma2} & = &\sqrt{\frac{1}{2}}(\alpha \beta \alpha-\beta \alpha \alpha)\nonumber\\
    \chi_{\frac{1}{2},-\frac{1}{2}}^{\sigma1} & = &\sqrt{\frac{1}{6}}(\alpha \beta \beta+\beta \alpha \beta-2 \beta \beta \alpha) \nonumber \\
    \chi_{\frac{1}{2},-\frac{1}{2}}^{\sigma2} & = &\sqrt{\frac{1}{2}}(\alpha \beta \beta-\beta \alpha \beta) \nonumber
\end{eqnarray}

The flavor wave functions of the $q^{3}$ cluster $\chi_{I, I_z}^{f}$ ($I$ and $I_z$ are the isospin quantum number and the third component, respectively) are as follows. Here, both the light and heavy quarks are considered as identical particles with the $SU(4)$ extension.
\begin{eqnarray}
    \chi_{0,0}^{f1}  & =& \frac{1}{2}(usd+sud-sdu-dsu) \nonumber \\   
    \chi_{0,0}^{f2}  & = &\sqrt{\frac{1}{12}}(2uds-2dsu+sdu+usd-sud-dsu) \nonumber \\
    \chi_{0,0}^{f3}  & = &\frac{1}{2}(ucd+cud-cdu-dcu) \nonumber \\  
    \chi_{0,0}^{f4}  & = &\sqrt{\frac{1}{12}}(2ucd-2dcu+cdu+ucd-cud-dcu) \nonumber \\
    \chi_{0,0}^{f5}  & = &\sqrt{\frac{1}{6}}(2ssc-scs-css) \nonumber \\  
    \chi_{0,0}^{f6}  & = &\sqrt{\frac{1}{2}}(scs-css) \nonumber
\end{eqnarray}
\begin{eqnarray}
    \chi_{0,0}^{f7}  & = &\sqrt{\frac{1}{3}}(ssc+scs+css) \nonumber \\   
    \chi_{0,0}^{f8}  & = &sss          \nonumber  \\
    \chi_{\frac{1}{2},-\frac{1}{2}}^{f1}  & =&\frac{1}{2}(dcs+cds-csd-scd) \nonumber \\ 
    \chi_{\frac{1}{2},-\frac{1}{2}}^{f2}  & =&\sqrt{\frac{1}{12}}(2dsc-2sdc+csd+dcs-cds-scd) \nonumber\\
    \chi_{\frac{1}{2},-\frac{1}{2}}^{f3}  & =&\sqrt{\frac{1}{6}}(udd+dud-2ddu) \nonumber \\  
    \chi_{\frac{1}{2},-\frac{1}{2}}^{f4}  & =&\sqrt{\frac{1}{2}}(udd-dud) \nonumber \\
    \chi_{\frac{1}{2},-\frac{1}{2}}^{f5}  & =&\sqrt{\frac{1}{12}}(2dsc+2sdc-csd-dcs-cds-scd) \nonumber \\
    \chi_{\frac{1}{2},-\frac{1}{2}}^{f6}  & =&\frac{1}{2}(dcs+scd-csd-cds) \nonumber \\
    \chi_{\frac{1}{2},-\frac{1}{2}}^{f7}  & =&\sqrt{\frac{1}{6}}(dss+sds-2ssd) \nonumber \\  
    \chi_{\frac{1}{2},-\frac{1}{2}}^{f8}  & =&\sqrt{\frac{1}{2}}(dss-sds) \nonumber \\
    \chi_{\frac{1}{2},-\frac{1}{2}}^{f9}  & =&\sqrt{\frac{1}{6}}(dsc+sdc+csd+dcs+cds+scd) \nonumber\\  
    \chi_{\frac{1}{2},-\frac{1}{2}}^{f10}  & =&\sqrt{\frac{1}{3}}(dss+sds+ssd) \nonumber \\
    \chi_{\frac{1}{2},\frac{1}{2}}^{f1}  & =&\sqrt{\frac{1}{6}}(2uud-udu-duu) \nonumber \\  
    \chi_{\frac{1}{2},\frac{1}{2}}^{f2}  & =&\sqrt{\frac{1}{2}}(udu-duu) \nonumber \\
    \chi_{\frac{1}{2},\frac{1}{2}}^{f3}  & =&\frac{1}{2}(ucs+cus-csu-scu) \nonumber \\
    \chi_{\frac{1}{2},\frac{1}{2}}^{f4}  & =&\sqrt{\frac{1}{12}}(2usc-2suc+csu+ucs-cus-scu) \nonumber\\
    \chi_{\frac{1}{2},\frac{1}{2}}^{f5}  & =&\sqrt{\frac{1}{12}}(2usc+2suc-csu-ucs-cus-scu) \nonumber\\
    \chi_{\frac{1}{2},\frac{1}{2}}^{f6}  & =&\frac{1}{2}(ucs+scu-csu-cus) \nonumber\\
        \chi_{\frac{1}{2},\frac{1}{2}}^{f7}  & =&\sqrt{\frac{1}{6}}(uss+sus-2ssu) \nonumber \\  
    \chi_{\frac{1}{2},\frac{1}{2}}^{f8}  & =&\sqrt{\frac{1}{2}}(uss-sus) \nonumber \\
    \chi_{\frac{1}{2},\frac{1}{2}}^{f9}  & =&\sqrt{\frac{1}{6}}(usc+suc+csu+ucs+cus+scu) \nonumber\\
    \chi_{\frac{1}{2},\frac{1}{2}}^{f10}  & =&\sqrt{\frac{1}{3}}(uss+sus+ssu)  \nonumber \\  
    \chi_{1,-1}^{f1}  & = &\sqrt{\frac{1}{6}}(2ddc-dcd-cdd) \nonumber \\  
    \chi_{1,-1}^{f2}  & = &\sqrt{\frac{1}{2}}(dcd-cdd) \nonumber \\
    \chi_{1,-1}^{f3}  & = &\sqrt{\frac{1}{6}}(2dds-dsd-sdd) \nonumber \\  
    \chi_{1,-1}^{f4}  & = &\sqrt{\frac{1}{2}}(dsd-sdd) \nonumber
\end{eqnarray}
\begin{eqnarray}
    \chi_{1,-1}^{f5}  & = &\sqrt{\frac{1}{3}}(ddc+dcd+cdd) \nonumber \\  
    \chi_{1,-1}^{f6}  & = &\sqrt{\frac{1}{3}}(dds+dsd+sdd) \nonumber  \\
    \chi_{1,0}^{f1}  & = &\sqrt{\frac{1}{12}}(2uds+2dus-sdu-usd-sud-dsu) \nonumber \\  
    \chi_{1,0}^{f2}  & = &\frac{1}{2}(usd+dsu-sdu-sud) \nonumber\\
    \chi_{1,0}^{f3}  & = &\sqrt{\frac{1}{12}}(2udc+2duc-cdu-ucd-cud-dcu) \nonumber \\  
    \chi_{1,0}^{f4}  & = &\frac{1}{2}(ucd+dcu-cdu-cud) \nonumber \\
    \chi_{1,0}^{f5}  & = &\sqrt{\frac{1}{6}}(udc+duc+cdu+ucd+cud+dcu) \nonumber \\  
    \chi_{1,0}^{f6}  & = &\sqrt{\frac{1}{6}}(uds+dus+sdu+usd+sud+dsu) \nonumber \\  
    \chi_{1,1}^{f1}  & = &\sqrt{\frac{1}{6}}(2uus-usu-suu) \nonumber \\  
    \chi_{1,1}^{f2}  & = &\sqrt{\frac{1}{2}}(usu-suu) \nonumber \\
    \chi_{1,1}^{f3}  & = &\sqrt{\frac{1}{6}}(2uuc-ucu-cuu) \nonumber \\  
    \chi_{1,1}^{f4}  & = &\sqrt{\frac{1}{2}}(ucu-cuu) \nonumber \\
    \chi_{1,1}^{f5}  & = &\sqrt{\frac{1}{3}}(uuc+ucu+cuu) \nonumber \\  
    \chi_{1,1}^{f6}  & = &\sqrt{\frac{1}{3}}(uus+usu+suu) \nonumber 
\end{eqnarray}

The color wave function of a color-singlet $q^{3}$ cluster is:
\begin{align}
	\chi^{c} =& \sqrt{\frac{1}{6}}(r g b-r b g+g b r-g r b+b r g-b g r) \nonumber
\end{align}

The total flavor-spin-color wave function of the dibaryon system can be acquired by substituting the wave functions of the flavor, the spin, and the color parts according to the given quantum number of the system, and the total flavor-spin-color wave function for each channel is shown as follows. $\phi_{I_{z},s_{z} }^{B}$ represents the wave function of the $q^{3}$ cluster ($I_{z}$ and $s_{z}$ are the third component of the isospin and spin quantum numbers, $B$ is the corresponding baryon). Then we couple the two baryon wave functions by Clebsch-Gordan coefficients according to the total quantum number requirement, and we can obtain the total wave functions. 
There are three channels for the $C=2,S=0$ system:
\begin{align}
\left | N\Xi_{cc}  \right \rangle =&\sqrt{\frac{1}{2} } \left [ \phi _{\frac{1}{2},\frac{1}{2}  }^{p} \phi _{-\frac{1}{2} ,\frac{1}{2} }^{\Xi _{cc} } -\phi _{-\frac{1}{2},\frac{1}{2}  }^{n} \phi _{\frac{1}{2} ,\frac{1}{2} }^{\Xi _{cc} }\right ] \nonumber \\
\left | N\Xi _{cc}^{\ast }  \right \rangle =&\sqrt{\frac{3}{8} } \left [ \phi _{-\frac{1}{2},\frac{3}{2}  }^{\Xi _{cc}^{\ast }} \phi _{-\frac{1}{2} ,-\frac{1}{2} }^{p }-\phi _{\frac{1}{2},\frac{3}{2}  }^{\Xi _{cc}^{\ast }} \phi _{-\frac{1}{2} ,-\frac{1}{2} }^{n}\right ] \nonumber \\
&-\sqrt{\frac{1}{8} }\left [\phi _{-\frac{1}{2},\frac{1}{2}  }^{\Xi _{cc}^{\ast }} \phi _{\frac{1}{2} ,\frac{1}{2} }^{p }-\phi _{\frac{1}{2},\frac{1}{2}  }^{\Xi _{cc}^{\ast }} \phi _{-\frac{1}{2} ,\frac{1}{2} }^{n }\right ]  \nonumber \\
\left |\Sigma_{c}^{\ast}\Sigma_{c}\right \rangle =&\frac{1}{2}\left [ \phi_{1,\frac{3}{2}}^{\Sigma_{c}^{\ast}}\phi_{-1,-\frac{1}{2}}^{\Sigma_{c}}
-\phi_{0,\frac{3}{2}}^{\Sigma_{c}^{\ast}}\phi_{0,-\frac{1}{2}}^{\Sigma_{c}}
+\phi_{-1,\frac{3}{2}}^{\Sigma_{c}^{\ast}}\phi_{1,-\frac{1}{2}}^{\Sigma_{c}} \right ] \nonumber \\
&-\sqrt{\frac{1}{12} }\left [ \phi_{1,\frac{1}{2}}^{\Sigma_{c}^{\ast}}\phi_{-1,\frac{1}{2}}^{\Sigma_{c}}
-\phi_{0,\frac{1}{2}}^{\Sigma_{c}^{\ast}}\phi_{0,\frac{1}{2}}^{\Sigma_{c}}
+\phi_{-1,\frac{1}{2}}^{\Sigma_{c}^{\ast}}\phi_{1,\frac{1}{2}}^{\Sigma_{c}} \right ] \nonumber
\end{align}

nine channels for the $C=2,S=-2$ system:
\begin{align}
\left |\Lambda \Omega_{cc}\right \rangle=&\phi_{0,\frac{1}{2}}^{\Lambda }\phi_{0,\frac{1}{2}}^{\Omega_{cc}}\nonumber \\
\left |\Lambda_{c} \Omega_{c}\right \rangle=&\phi_{0,\frac{1}{2}}^{\Lambda_{c} }\phi_{0,\frac{1}{2}}^{\Omega_{c}}\nonumber \\
\left |\Lambda\Omega_{cc}^{\ast}\right \rangle=&\sqrt{\frac{3}{4}}\phi_{0,-\frac{1}{2}}^{\Lambda}\phi_{0,\frac{3}{2}}^{\Omega_{cc}^{\ast }}
+\sqrt{\frac{1}{4}}\phi_{0,\frac{1}{2}}^{\Lambda}\phi_{0,\frac{1}{2}}^{\Omega_{cc}^{\ast }}  \nonumber \\
\left |\Xi \Xi_{cc} \right \rangle=&\sqrt{\frac{1}{2}}\left [\phi_{\frac{1}{2},\frac{1}{2}}^{\Xi }\phi_{-\frac{1}{2},\frac{1}{2}}^{\Xi_{cc}}
-\phi_{-\frac{1}{2},\frac{1}{2}}^{\Xi }\phi_{\frac{1}{2},\frac{1}{2}}^{\Xi_{cc}} \right ]\nonumber \\
\left |\Xi_{c} \Xi_{c} \right \rangle=&\sqrt{\frac{1}{2}}\left [\phi_{\frac{1}{2},\frac{1}{2}}^{\Xi_{c} }\phi_{-\frac{1}{2},\frac{1}{2}}^{\Xi_{c}}
-\phi_{-\frac{1}{2},\frac{1}{2}}^{\Xi_{c} }\phi_{\frac{1}{2},\frac{1}{2}}^{\Xi_{c}} \right ] \nonumber \\
\left |\Xi_{c} \Xi_{c}^{\prime } \right \rangle = &\sqrt{\frac{1}{2}}\left [\phi_{\frac{1}{2},\frac{1}{2}}^{\Xi_{c} }\phi_{-\frac{1}{2},\frac{1}{2}}^{\Xi_{c}^{\prime }}
-\phi_{-\frac{1}{2},\frac{1}{2}}^{\Xi_{c} }\phi_{\frac{1}{2},\frac{1}{2}}^{\Xi_{c}^{\prime }} \right ]\nonumber \\
\left |\Xi_{c}^{\prime } \Xi_{c}^{\prime } \right \rangle =& \sqrt{\frac{1}{2}}\left [\phi_{\frac{1}{2},\frac{1}{2}}^{\Xi_{c}^{\prime } }\phi_{-\frac{1}{2},\frac{1}{2}}^{\Xi_{c}^{\prime }}
-\phi_{-\frac{1}{2},\frac{1}{2}}^{\Xi_{c}^{\prime } }\phi_{\frac{1}{2},\frac{1}{2}}^{\Xi_{c}^{\prime }} \right ]\nonumber \\
\left |\Xi^{\ast } \Xi_{cc}\right \rangle =& \sqrt{\frac{3}{8} }\left [\phi_{\frac{1}{2},\frac{3}{2}}^{\Xi^{\ast } }\phi_{-\frac{1}{2},-\frac{1}{2}}^{\Xi_{cc}}
-\phi_{-\frac{1}{2},\frac{3}{2}}^{\Xi^{\ast }} \phi_{\frac{1}{2},-\frac{1}{2}}^{\Xi_{cc}} \right ]\nonumber \\
&-\sqrt{\frac{1}{8} }\left [\phi_{\frac{1}{2},\frac{1}{2}}^{\Xi^{\ast } }\phi_{-\frac{1}{2},\frac{1}{2}}^{\Xi_{cc}}
-\phi_{-\frac{1}{2},\frac{1}{2}}^{\Xi^{\ast }} \phi_{\frac{1}{2},\frac{1}{2}}^{\Xi_{cc}} \right ]\nonumber \\
\left |\Xi_{c}^{\ast } \Xi_{c}\right \rangle = &\sqrt{\frac{3}{8} }\left [\phi_{\frac{1}{2},\frac{3}{2}}^{\Xi_{c}^{\ast } }\phi_{-\frac{1}{2},-\frac{1}{2}}^{\Xi_{c}}
-\phi_{-\frac{1}{2},\frac{3}{2}}^{\Xi_{c}^{\ast }} \phi_{\frac{1}{2},-\frac{1}{2}}^{\Xi_{c}} \right ]\nonumber \\
&-\sqrt{\frac{1}{8} }\left [\phi_{\frac{1}{2},\frac{1}{2}}^{\Xi_{c}^{\ast } }\phi_{-\frac{1}{2},\frac{1}{2}}^{\Xi_{c}}
-\phi_{-\frac{1}{2},\frac{1}{2}}^{\Xi_{c}^{\ast }} \phi_{\frac{1}{2},\frac{1}{2}}^{\Xi_{c}} \right ]\nonumber \\
\left |\Xi_{c}^{\ast } \Xi_{c}^{\prime }\right \rangle =& \sqrt{\frac{3}{8} }\left [\phi_{\frac{1}{2},\frac{3}{2}}^{\Xi_{c}^{\ast } }\phi_{-\frac{1}{2},-\frac{1}{2}}^{\Xi_{c}^{\prime }}
-\phi_{-\frac{1}{2},\frac{3}{2}}^{\Xi_{c}^{\ast }} \phi_{\frac{1}{2},-\frac{1}{2}}^{\Xi_{c}^{\prime }} \right ]\nonumber \\
&-\sqrt{\frac{1}{8} }\left [\phi_{\frac{1}{2},\frac{1}{2}}^{\Xi_{c}^{\ast } }\phi_{-\frac{1}{2},\frac{1}{2}}^{\Xi_{c}^{\prime }}
-\phi_{-\frac{1}{2},\frac{1}{2}}^{\Xi_{c}^{\ast }} \phi_{\frac{1}{2},\frac{1}{2}}^{\Xi_{c}^{\prime }} \right ]\nonumber \\
\left |\Xi_{cc}^{\ast } \Xi\right \rangle = &\sqrt{\frac{3}{8} }\left [\phi_{\frac{1}{2},\frac{3}{2}}^{\Xi_{cc}^{\ast } }\phi_{-\frac{1}{2},-\frac{1}{2}}^{\Xi}
-\phi_{-\frac{1}{2},\frac{3}{2}}^{\Xi_{cc}^{\ast }} \phi_{\frac{1}{2},-\frac{1}{2}}^{\Xi} \right ]\nonumber \\
&-\sqrt{\frac{1}{8} }\left [\phi_{\frac{1}{2},\frac{1}{2}}^{\Xi_{cc}^{\ast } }\phi_{-\frac{1}{2},\frac{1}{2}}^{\Xi}
-\phi_{-\frac{1}{2},\frac{1}{2}}^{\Xi_{cc}^{\ast }} \phi_{\frac{1}{2},\frac{1}{2}}^{\Xi} \right ]\nonumber
\end{align}
\begin{align}
\left |\Xi^{\ast } \Xi_{cc}^{\ast }\right \rangle = &\sqrt{\frac{3}{20} }\left [\phi_{\frac{1}{2},\frac{3}{2}}^{\Xi^{\ast } }\phi_{-\frac{1}{2},-\frac{1}{2}}^{\Xi_{cc}^{\ast }}
-\phi_{-\frac{1}{2},\frac{3}{2}}^{\Xi^{\ast }} \phi_{\frac{1}{2},-\frac{1}{2}}^{\Xi_{cc}^{\ast }} \right. \nonumber \\
&\left.+\phi_{\frac{1}{2},-\frac{1}{2}}^{\Xi^{\ast }} \phi_{-\frac{1}{2},\frac{3}{2}}^{\Xi_{cc}^{\ast }}
-\phi_{-\frac{1}{2},-\frac{1}{2}}^{\Xi^{\ast }} \phi_{\frac{1}{2},\frac{3}{2}}^{\Xi_{cc}^{\ast }}\right ]\nonumber \\
&-\sqrt{\frac{1}{5} }\left [\phi_{\frac{1}{2},\frac{1}{2}}^{\Xi^{\ast } }\phi_{-\frac{1}{2},\frac{1}{2}}^{\Xi_{cc}^{\ast}}
-\phi_{-\frac{1}{2},\frac{1}{2}}^{\Xi^{\ast }} \phi_{\frac{1}{2},\frac{1}{2}}^{\Xi_{cc}^{\ast }} \right ]\nonumber \\
\left |\Xi_{c}^{\ast } \Xi_{c}^{\ast }\right \rangle =& \sqrt{\frac{3}{20} }\left [\phi_{\frac{1}{2},\frac{3}{2}}^{\Xi_{c}^{\ast } }\phi_{-\frac{1}{2},-\frac{1}{2}}^{\Xi_{c}^{\ast }}
-\phi_{-\frac{1}{2},\frac{3}{2}}^{\Xi_{c}^{\ast }} \phi_{\frac{1}{2},-\frac{1}{2}}^{\Xi_{c}^{\ast }}\right. \nonumber \\
&\left. +\phi_{\frac{1}{2},-\frac{1}{2}}^{\Xi_{c}^{\ast }} \phi_{-\frac{1}{2},\frac{3}{2}}^{\Xi_{c}^{\ast }}
-\phi_{-\frac{1}{2},-\frac{1}{2}}^{\Xi_{c}^{\ast }} \phi_{\frac{1}{2},\frac{3}{2}}^{\Xi_{c}^{\ast }}\right ]\nonumber \\
&-\sqrt{\frac{1}{5} }\left [\phi_{\frac{1}{2},\frac{1}{2}}^{\Xi_{c}^{\ast } }\phi_{-\frac{1}{2},\frac{1}{2}}^{\Xi_{c}^{\ast}}
-\phi_{-\frac{1}{2},\frac{1}{2}}^{\Xi_{c}^{\ast }} \phi_{\frac{1}{2},\frac{1}{2}}^{\Xi_{c}^{\ast }} \right ]\nonumber
\end{align}

and three channels for the $C=2,S=-4$ system:
\begin{align}
\left |\Omega \Omega_{cc} \right \rangle =& \sqrt{\frac{3}{4} }\phi_{0,\frac{3}{2}}^{\Omega  }\phi_{0,-\frac{1}{2}}^{\Omega _{cc}}
-\sqrt{\frac{1}{4} } \phi_{0,\frac{1}{2}}^{\Omega } \phi_{0,\frac{1}{2}}^{\Omega_{cc}} \nonumber \\
\left |\Omega_{c}^{\ast } \Omega_{c} \right \rangle = &\sqrt{\frac{3}{4} }\phi_{0,\frac{3}{2}}^{\Omega_{c}^{\ast } }\phi_{0,-\frac{1}{2}}^{\Omega _{c}}
-\sqrt{\frac{1}{4} } \phi_{0,\frac{1}{2}}^{\Omega_{c}^{\ast } } \phi_{0,\frac{1}{2}}^{\Omega_{c}}\nonumber \\
\left |\Omega \Omega_{cc}^{\ast } \right \rangle = &\sqrt{\frac{3}{10} }\phi_{0,\frac{3}{2}}^{\Omega }\phi_{0,-\frac{1}{2}}^{\Omega _{cc}^{\ast }} \nonumber \\
&-\sqrt{\frac{2}{5} } \phi_{0,\frac{1}{2}}^{\Omega } \phi_{0,\frac{1}{2}}^{\Omega_{cc}^{\ast }}+ \sqrt{\frac{3}{10} }\phi_{0,-\frac{1}{2}}^{\Omega }\phi_{0,\frac{3}{2}}^{\Omega _{cc}^{\ast }} \nonumber
\end{align}

where the expression of $\phi_{I_{z},s_{z} }^{B}$ is shown as follows:
\begin{align}
\phi _{0,\frac{1}{2}}^{\Lambda}=& \sqrt{\frac{1}{2} }\left (  \chi _{0,0}^{f1}\chi _{\frac{1}{2},\frac{1}{2}}^{\sigma 1} +\chi _{0,0}^{f2}\chi _{\frac{1}{2},\frac{1}{2}}^{\sigma 2} \right )\chi ^{c}\nonumber \\
\phi _{0,\frac{1}{2}}^{\Lambda_{c}}=& \sqrt{\frac{1}{2}}\left (  \chi _{0,0}^{f3}\chi _{\frac{1}{2},\frac{1}{2}}^{\sigma 1} +\chi _{0,0}^{f4}\chi _{\frac{1}{2},\frac{1}{2}}^{\sigma 2} \right )\chi ^{c}\nonumber \\
\phi _{\frac{1}{2},\frac{1}{2}}^{p}=& \sqrt{\frac{1}{2}}\left (  \chi _{\frac{1}{2},\frac{1}{2}}^{f1}\chi _{\frac{1}{2},\frac{1}{2}}^{\sigma 1} +\chi _{\frac{1}{2},\frac{1}{2}}^{f2}\chi _{\frac{1}{2},\frac{1}{2}}^{\sigma 2} \right )\chi ^{c}\nonumber \\
\phi _{-\frac{1}{2},\frac{1}{2}}^{\Xi_{c}}=& \sqrt{\frac{1}{2}}\left (  \chi _{\frac{1}{2},-\frac{1}{2}}^{f1}\chi _{\frac{1}{2},\frac{1}{2}}^{\sigma 1} +\chi _{\frac{1}{2},-\frac{1}{2}}^{f2}\chi _{\frac{1}{2},\frac{1}{2}}^{\sigma 2} \right )\chi ^{c}\nonumber \\
\phi _{-\frac{1}{2},\frac{1}{2}}^{n}=& \sqrt{\frac{1}{2}}\left (  \chi _{\frac{1}{2},-\frac{1}{2}}^{f3}\chi _{\frac{1}{2},\frac{1}{2}}^{\sigma 1} +\chi _{\frac{1}{2},-\frac{1}{2}}^{f4}\chi _{\frac{1}{2},\frac{1}{2}}^{\sigma 2} \right )\chi ^{c}\nonumber \\
\phi _{\frac{1}{2},\frac{1}{2}}^{\Xi_{c}}=& \sqrt{\frac{1}{2}}\left (  \chi _{\frac{1}{2},\frac{1}{2}}^{f3}\chi _{\frac{1}{2},\frac{1}{2}}^{\sigma 1} +\chi _{\frac{1}{2},\frac{1}{2}}^{f4}\chi _{\frac{1}{2},\frac{1}{2}}^{\sigma 2} \right )\chi ^{c}\nonumber \\
\phi _{-\frac{1}{2},\frac{1}{2}}^{\Xi_{c}^{\prime}}=& \sqrt{\frac{1}{2}}\left (  \chi _{\frac{1}{2},-\frac{1}{2}}^{f5}\chi _{\frac{1}{2},\frac{1}{2}}^{\sigma 1} +\chi _{\frac{1}{2},-\frac{1}{2}}^{f6}\chi _{\frac{1}{2},\frac{1}{2}}^{\sigma 2} \right )\chi ^{c}\nonumber \\
\phi _{\frac{1}{2},\frac{1}{2}}^{\Xi_{c}^{\prime}}=& \sqrt{\frac{1}{2}}\left (  \chi _{\frac{1}{2},\frac{1}{2}}^{f5}\chi _{\frac{1}{2},\frac{1}{2}}^{\sigma 1} +\chi _{\frac{1}{2},\frac{1}{2}}^{f6}\chi _{\frac{1}{2},\frac{1}{2}}^{\sigma 2} \right )\chi ^{c}\nonumber
\end{align}
\begin{align}
\phi _{1,\frac{1}{2}}^{\Sigma}=& \sqrt{\frac{1}{2}}\left (  \chi _{1,1}^{f1}\chi _{\frac{1}{2},\frac{1}{2}}^{\sigma 1} +\chi _{1,1}^{f2}\chi _{\frac{1}{2},\frac{1}{2}}^{\sigma 2} \right )\chi ^{c}\nonumber \\
\phi _{-1,\frac{1}{2}}^{\Sigma_{c}}=& \sqrt{\frac{1}{2}}\left (  \chi _{1,-1}^{f1}\chi _{\frac{1}{2},\frac{1}{2}}^{\sigma 1} +\chi _{1,-1}^{f2}\chi _{\frac{1}{2},\frac{1}{2}}^{\sigma 2} \right )\chi ^{c}\nonumber \\
\phi _{0,\frac{1}{2}}^{\Sigma}=& \sqrt{\frac{1}{2}}\left (  \chi _{1,0}^{f1}\chi _{\frac{1}{2},\frac{1}{2}}^{\sigma 1} +\chi _{1,0}^{f2}\chi _{\frac{1}{2},\frac{1}{2}}^{\sigma 2} \right )\chi ^{c}\nonumber \\
\phi _{0,\frac{1}{2}}^{\Sigma_{c}}=& \sqrt{\frac{1}{2}}\left (  \chi _{1,0}^{f3}\chi _{\frac{1}{2},\frac{1}{2}}^{\sigma 1} +\chi _{1,0}^{f4}\chi _{\frac{1}{2},\frac{1}{2}}^{\sigma 2} \right )\chi ^{c}\nonumber \\
\phi _{-1,\frac{1}{2}}^{\Sigma}=& \sqrt{\frac{1}{2}}\left (  \chi _{1,-1}^{f3}\chi _{\frac{1}{2},\frac{1}{2}}^{\sigma 1} +\chi _{1,-1}^{f4}\chi _{\frac{1}{2},\frac{1}{2}}^{\sigma 2} \right )\chi ^{c}\nonumber \\
\phi _{1,\frac{1}{2}}^{\Sigma_{c}}=& \sqrt{\frac{1}{2}}\left (  \chi _{1,1}^{f3}\chi _{\frac{1}{2},\frac{1}{2}}^{\sigma 1} +\chi _{1,1}^{f4}\chi _{\frac{1}{2},\frac{1}{2}}^{\sigma 2} \right )\chi ^{c}\nonumber \\
\phi _{0,-\frac{1}{2}}^{\Sigma}=& \sqrt{\frac{1}{2}}\left (  \chi _{1,0}^{f1}\chi _{\frac{1}{2},-\frac{1}{2}}^{\sigma 1} +\chi _{1,0}^{f2}\chi _{\frac{1}{2},-\frac{1}{2}}^{\sigma 2} \right )\chi ^{c}\nonumber \\
\phi _{1,-\frac{1}{2}}^{\Sigma}=& \sqrt{\frac{1}{2}}\left (  \chi _{1,1}^{f1}\chi _{\frac{1}{2},-\frac{1}{2}}^{\sigma 1} +\chi _{1,1}^{f2}\chi _{\frac{1}{2},-\frac{1}{2}}^{\sigma 2} \right )\chi ^{c}\nonumber \\
\phi _{-1,-\frac{1}{2}}^{\Sigma}=& \sqrt{\frac{1}{2}}\left (  \chi _{1,-1}^{f3}\chi _{\frac{1}{2},-\frac{1}{2}}^{\sigma 1} +\chi _{1,-1}^{f4}\chi _{\frac{1}{2},-\frac{1}{2}}^{\sigma 2} \right )\chi ^{c}\nonumber \\
\phi _{0,\frac{1}{2}}^{\Omega_{c}}=& \sqrt{\frac{1}{2} }\left (  \chi _{0,0}^{f5}\chi _{\frac{1}{2},\frac{1}{2}}^{\sigma 1} +\chi _{0,0}^{f6}\chi _{\frac{1}{2},\frac{1}{2}}^{\sigma 2} \right )\chi ^{c}\nonumber \\
\phi _{0,-\frac{1}{2}}^{\Lambda}=& \sqrt{\frac{1}{2} }\left (  \chi _{0,0}^{f1}\chi _{\frac{1}{2},-\frac{1}{2}}^{\sigma 1} +\chi _{0,0}^{f2}\chi _{\frac{1}{2},-\frac{1}{2}}^{\sigma 2} \right )\chi ^{c}\nonumber \\
\phi _{\frac{1}{2},\frac{1}{2}}^{\Xi}=& \sqrt{\frac{1}{2} }\left (  \chi _{\frac{1}{2},\frac{1}{2}}^{f7}\chi _{\frac{1}{2},\frac{1}{2}}^{\sigma 1} +\chi _{\frac{1}{2},\frac{1}{2}}^{f8}\chi _{\frac{1}{2},\frac{1}{2}}^{\sigma 2} \right )\chi ^{c}\nonumber \\
\phi _{-\frac{1}{2},\frac{1}{2}}^{\Xi}=& \sqrt{\frac{1}{2} }\left (  \chi _{\frac{1}{2},-\frac{1}{2}}^{f7}\chi _{\frac{1}{2},\frac{1}{2}}^{\sigma 1} +\chi _{\frac{1}{2},-\frac{1}{2}}^{f8}\chi _{\frac{1}{2},\frac{1}{2}}^{\sigma 2} \right )\chi ^{c}\nonumber \\
\phi _{-\frac{1}{2},-\frac{1}{2}}^{\Xi}=& \sqrt{\frac{1}{2} }\left (  \chi _{\frac{1}{2},-\frac{1}{2}}^{f7}\chi _{\frac{1}{2},-\frac{1}{2}}^{\sigma 1} +\chi _{\frac{1}{2},-\frac{1}{2}}^{f8}\chi _{\frac{1}{2},-\frac{1}{2}}^{\sigma 2} \right )\chi ^{c}\nonumber \\
\phi _{\frac{1}{2},-\frac{1}{2}}^{\Xi}=& \sqrt{\frac{1}{2} }\left (  \chi _{\frac{1}{2},\frac{1}{2}}^{f7}\chi _{\frac{1}{2},-\frac{1}{2}}^{\sigma 1} +\chi _{\frac{1}{2},\frac{1}{2}}^{f8}\chi _{\frac{1}{2},-\frac{1}{2}}^{\sigma 2} \right )\chi ^{c}\nonumber \\
\phi _{-\frac{1}{2},-\frac{1}{2}}^{\Xi_{c}^{\prime}}=& \sqrt{\frac{1}{2} }\left (  \chi _{\frac{1}{2},-\frac{1}{2}}^{f5}\chi _{\frac{1}{2},-\frac{1}{2}}^{\sigma 1} +\chi _{\frac{1}{2},-\frac{1}{2}}^{f6}\chi _{\frac{1}{2},-\frac{1}{2}}^{\sigma 2} \right )\chi ^{c}\nonumber \\
\phi _{\frac{1}{2},-\frac{1}{2}}^{\Xi_{c}^{\prime}}=& \sqrt{\frac{1}{2} }\left (  \chi _{\frac{1}{2},\frac{1}{2}}^{f5}\chi _{\frac{1}{2},-\frac{1}{2}}^{\sigma 1} +\chi _{\frac{1}{2},\frac{1}{2}}^{f6}\chi _{\frac{1}{2},-\frac{1}{2}}^{\sigma 2} \right )\chi ^{c}\nonumber \\
\phi _{0,-\frac{1}{2}}^{\Omega_{c}}=& \sqrt{\frac{1}{2} }\left (  \chi _{0,0}^{f5}\chi _{\frac{1}{2},-\frac{1}{2}}^{\sigma 1} +\chi _{0,0}^{f6}\chi _{\frac{1}{2},-\frac{1}{2}}^{\sigma 2} \right )\chi ^{c}\nonumber \\
\phi _{-\frac{1}{2},\frac{1}{2}}^{\Xi_{cc}}=& \sqrt{\frac{1}{2} }\left (  \chi _{\frac{1}{2},-\frac{1}{2}}^{f11}\chi _{\frac{1}{2},\frac{1}{2}}^{\sigma 1} +\chi _{\frac{1}{2},-\frac{1}{2}}^{f12}\chi _{\frac{1}{2},\frac{1}{2}}^{\sigma 2} \right )\chi ^{c}\nonumber \\
\phi _{-\frac{1}{2},-\frac{1}{2}}^{\Xi_{cc}}=& \sqrt{\frac{1}{2} }\left (  \chi _{\frac{1}{2},-\frac{1}{2}}^{f11}\chi _{\frac{1}{2},-\frac{1}{2}}^{\sigma 1} +\chi _{\frac{1}{2},-\frac{1}{2}}^{f12}\chi _{\frac{1}{2},-\frac{1}{2}}^{\sigma 2} \right )\chi ^{c} \nonumber \\
\phi _{\frac{1}{2},\frac{1}{2}}^{\Xi_{cc}}=& \sqrt{\frac{1}{2} }\left (  \chi _{\frac{1}{2},\frac{1}{2}}^{f11}\chi _{\frac{1}{2},\frac{1}{2}}^{\sigma 1} +\chi _{\frac{1}{2},\frac{1}{2}}^{f12}\chi _{\frac{1}{2},\frac{1}{2}}^{\sigma 2} \right )\chi ^{c}\nonumber \\
\phi _{\frac{1}{2},-\frac{1}{2}}^{\Xi_{cc}}=& \sqrt{\frac{1}{2} }\left (  \chi _{\frac{1}{2},\frac{1}{2}}^{f11}\chi _{\frac{1}{2},-\frac{1}{2}}^{\sigma 1} +\chi _{\frac{1}{2},\frac{1}{2}}^{f12}\chi _{\frac{1}{2},-\frac{1}{2}}^{\sigma 2} \right )\chi ^{c}  \nonumber \\
\phi _{0,\frac{1}{2}}^{\Omega_{cc}}=& \sqrt{\frac{1}{2} }\left (  \chi _{0,0}^{f9}\chi _{\frac{1}{2},\frac{1}{2}}^{\sigma 1} +\chi _{0,0}^{f10}\chi _{\frac{1}{2},\frac{1}{2}}^{\sigma 2} \right )\chi ^{c}\nonumber \\
\phi _{0,-\frac{1}{2}}^{\Omega_{cc}}=& \sqrt{\frac{1}{2} }\left (  \chi _{0,0}^{f9}\chi _{\frac{1}{2},-\frac{1}{2}}^{\sigma 1} +\chi _{0,0}^{f10}\chi _{\frac{1}{2},-\frac{1}{2}}^{\sigma 2} \right )\chi ^{c}  \nonumber
\end{align}
\begin{align}
\phi_{0,\frac{3}{2}}^{\Sigma_{c}^*}=&\chi_{1,0}^{f5}\chi_
{\frac{3}{2},\frac{3}{2}}^{\sigma}\chi^{c} ~~~~~~~~~~~~\phi_{-1,\frac{3}{2}}^{\Sigma_{c}^*}=\chi_{1,-1}^{f5}
\chi_{\frac{3}{2},\frac{3}{2}}^{\sigma}\chi^{c} \nonumber \\
\phi_{1,\frac{3}{2}}^{\Sigma_{c}^*}=&\chi_{1,1}^{f5}\chi_
{\frac{3}{2},\frac{3}{2}}^{\sigma}\chi^{c}~~~~~~~~~~~~
\phi_{0,\frac{1}{2}}^{\Sigma_{c}^*}=\chi_{1,0}^{f5}
\chi_{\frac{3}{2},\frac{1}{2}}^{\sigma}\chi^{c}\nonumber \\
\phi_{-1,\frac{1}{2}}^{\Sigma_{c}^*}=&\chi_{1,-1}^{f5}\chi_
{\frac{3}{2},\frac{1}{2}}^{\sigma}\chi^{c} ~~~~~~~~~~~~\phi_{1,\frac{1}{2}}^{\Sigma_{c}^*}=\chi_{1,1}^{f5}
\chi_{\frac{3}{2},\frac{1}{2}}^{\sigma}\chi^{c}\nonumber \\
\phi_{1,\frac{3}{2}}^{\Sigma^*}=&\chi_{1,1}^{f6}\chi_
{\frac{3}{2},\frac{3}{2}}^{\sigma}\chi^{c} ~~~~~~~~~~~~\phi_{0,\frac{3}{2}}^{\Sigma^*}=\chi_{1,0}^{f6}
\chi_{\frac{3}{2},\frac{3}{2}}^{\sigma}\chi^{c}\nonumber \\
\phi_{-1,\frac{3}{2}}^{\Sigma^*}=&\chi_{1,-1}^{f6}\chi_
{\frac{3}{2},\frac{3}{2}}^{\sigma}\chi^{c} ~~~~~~~~~~~~
\phi_{1,\frac{1}{2}}^{\Sigma^*}=\chi_{1,1}^{f6}
\chi_{\frac{3}{2},\frac{1}{2}}^{\sigma}\chi^{c}\nonumber \\
\phi_{0,\frac{1}{2}}^{\Sigma^*}=&\chi_{1,0}^{f6}\chi_
{\frac{3}{2},\frac{1}{2}}^{\sigma}\chi^{c} ~~~~~~~~~~~~~~
\phi_{-1,\frac{1}{2}}^{\Sigma^*}=\chi_{1,-1}^{f6}
\chi_{\frac{3}{2},\frac{1}{2}}^{\sigma}\chi^{c}\nonumber \\
\phi_{-1,-\frac{1}{2}}^{\Sigma_{c}^*}=&\chi_{1,-1}^{f5}\chi_
{\frac{3}{2},-\frac{1}{2}}^{\sigma}\chi^{c} ~~~~~~~\phi_{0,-\frac{1}{2}}^{\Sigma_{c}^*}=\chi_{1,0}^{f5}
\chi_{\frac{3}{2},-\frac{1}{2}}^{\sigma}\chi^{c}\nonumber \\
\phi_{1,-\frac{1}{2}}^{\Sigma_{c}^*}=&\chi_{1,1}^{f5}\chi_
{\frac{3}{2},-\frac{1}{2}}^{\sigma}\chi^{c}~~~~~~~~~~~~
\phi_{1,-\frac{1}{2}}^{\Sigma^*}=\chi_{1,1}^{f6}
\chi_{\frac{3}{2},-\frac{1}{2}}^{\sigma}\chi^{c}\nonumber \\
\phi_{0,-\frac{1}{2}}^{\Sigma_{c}^*}=&\chi_{1,0}^{f6}\chi_
{\frac{3}{2},-\frac{1}{2}}^{\sigma}\chi^{c}~~~~~~~~~~~~
\phi_{-1,-\frac{1}{2}}^{\Sigma^*}=\chi_{1,-1}^{f6}
\chi_{\frac{3}{2},-\frac{1}{2}}^{\sigma}\chi^{c}\nonumber \\
\phi_{0,\frac{3}{2}}^{\Omega_{c}^*}=&\chi_{0,0}^{f7}\chi_
{\frac{3}{2},\frac{3}{2}}^{\sigma}\chi^{c} ~~~~~~~~~~~~~\phi_{0,\frac{1}{2}}^{\Omega_{c}^*}=\chi_{0,0}^{f7}
\chi_{\frac{3}{2},\frac{1}{2}}^{\sigma}\chi^{c}\nonumber \\
\phi_{0,\frac{1}{2}}^{\Omega}=&\chi_{0,0}^{f8}\chi_
{\frac{3}{2},\frac{1}{2}}^{\sigma}\chi^{c} ~~~~~~~~~~~~
\phi_{0,\frac{3}{2}}^{\Omega}=\chi_{0,0}^{f8}
\chi_{\frac{3}{2},\frac{3}{2}}^{\sigma}\chi^{c}\nonumber \\
\phi_{\frac{1}{2},\frac{3}{2}}^{\Xi_{c}^*}=&\chi_
{\frac{1}{2},\frac{1}{2}}^{f9}\chi_{\frac{3}{2},
\frac{3}{2}}^{\sigma}\chi^{c}~~~~~~~~~~~~
\phi_{-\frac{1}{2},\frac{3}{2}}^{\Xi_{c}^*}=\chi_
{\frac{1}{2},-\frac{1}{2}}^{f9}\chi_{\frac{3}{2},
\frac{3}{2}}^{\sigma}\chi^{c}\nonumber \\
\phi_{\frac{1}{2},\frac{1}{2}}^{\Xi_{c}^*}=&\chi_
{\frac{1}{2},\frac{1}{2}}^{f9}\chi_{\frac{3}{2},
\frac{1}{2}}^{\sigma}\chi^{c} ~~~~~~~~~~~~\phi_{-\frac{1}{2},\frac{1}{2}}^{\Xi_{c}^*}=\chi_
{\frac{1}{2},-\frac{1}{2}}^{f9}\chi_{\frac{3}{2},
\frac{1}{2}}^{\sigma}\chi^{c}\nonumber \\
\phi_{\frac{1}{2},\frac{3}{2}}^{\Xi^*}=&\chi_
{\frac{1}{2},\frac{1}{2}}^{f10}\chi_{\frac{3}{2},
\frac{3}{2}}^{\sigma}\chi^{c} ~~~~~~~~~~~~
\phi_{-\frac{1}{2},\frac{3}{2}}^{\Xi^*}=\chi_
{\frac{1}{2},-\frac{1}{2}}^{f10}\chi_{\frac{3}{2},
\frac{3}{2}}^{\sigma}\chi^{c}\nonumber \\
\phi_{\frac{1}{2},\frac{1}{2}}^{\Xi^*}=&\chi_
{\frac{1}{2},\frac{1}{2}}^{f10}\chi_{\frac{3}{2},
\frac{1}{2}}^{\sigma}\chi^{c} ~~~~~~~~~~~~
\phi_{-\frac{1}{2},\frac{1}{2}}^{\Xi^*}=\chi_
{\frac{1}{2},-\frac{1}{2}}^{f10}\chi_{\frac{3}{2},
\frac{1}{2}}^{\sigma}\chi^{c} \nonumber \\
\phi_{-\frac{1}{2},-\frac{1}{2}}^{\Xi_{c}^*}=&\chi_
{\frac{1}{2},-\frac{1}{2}}^{f9}\chi_{\frac{3}{2},
-\frac{1}{2}}^{\sigma}\chi^{c} ~~~~~~~~~~~~\phi_{\frac{1}{2},-\frac{1}{2}}^{\Xi_{c}^*}=\chi_
{\frac{1}{2},\frac{1}{2}}^{f9}\chi_{\frac{3}{2},
-\frac{1}{2}}^{\sigma}\chi^{c}\nonumber \\
\phi_{\frac{1}{2},-\frac{1}{2}}^{\Xi^*}=&\chi_
{\frac{1}{2},\frac{1}{2}}^{f10}\chi_{\frac{3}{2},
-\frac{1}{2}}^{\sigma}\chi^{c}~~~~~~~~~~~~
\phi_{-\frac{1}{2},-\frac{1}{2}}^{\Xi^*}=\chi_
{\frac{1}{2},-\frac{1}{2}}^{f10}\chi_{\frac{3}{2},
-\frac{1}{2}}^{\sigma}\chi^{c}\nonumber \\
\phi_{0,-\frac{1}{2}}^{\Omega}=&\chi_{0,0}^{f8}\chi_
{\frac{3}{2},-\frac{1}{2}}^{\sigma}\chi^{c}~~~~~~~~~~~~
\phi_{0,\frac{3}{2}}^{\Omega_{c}^*}=\chi_{0,0}^{f7}
\chi_{\frac{3}{2},-\frac{1}{2}}^{\sigma}\chi^{c} \nonumber\\
\phi_{-\frac{1}{2},\frac{3}{2}}^{\Xi_cc^*}=&\chi_
{\frac{1}{2},-\frac{1}{2}}^{f13}\chi_{\frac{3}{2},
\frac{3}{2}}^{\sigma}\chi^{c}~~~~~~~~~~~~
\phi_{-\frac{1}{2},\frac{1}{2}}^{\Xi_{cc}^*}=\chi_{\frac{1}{2},-\frac{1}{2}}^{f13}\chi_{\frac{3}{2},
\frac{1}{2}}^{\sigma}\chi^{c}\nonumber \\
\phi_{-\frac{1}{2},-\frac{1}{2}}^{\Xi_{cc}^*}=&\chi_{\frac{1}{2},-\frac{1}{2}}^{f13}\chi_{\frac{3}{2},
-\frac{1}{2}}^{\sigma}\chi^{c} ~~~~~~~~~~~~
\phi_{\frac{1}{2},\frac{3}{2}}^{\Xi_{cc}^*}=\chi_{\frac{1}{2},\frac{1}{2}}^{f13}\chi_{\frac{3}{2},
\frac{3}{2}}^{\sigma}\chi^{c} \nonumber \\
\phi_{\frac{1}{2},\frac{1}{2}}^{\Xi_{cc}^*}=&\chi_{\frac{1}{2},\frac{1}{2}}^{f13}\chi_{\frac{3}{2},
\frac{1}{2}}^{\sigma}\chi^{c}~~~~~~~~~~~~
\phi_{\frac{1}{2},-\frac{1}{2}}^{\Xi_{cc}^*}=\chi_{\frac{1}{2},\frac{1}{2}}^{f13}\chi_{\frac{3}{2},
-\frac{1}{2}}^{\sigma}\chi^{c} \nonumber\\
\phi_{0,\frac{3}{2}}^{\Omega_{cc}}=&\chi_{0,0}^{f11}\chi_
{\frac{3}{2},\frac{3}{2}}^{\sigma}\chi^{c}~~~~~~~~~~~~
\phi_{0,\frac{1}{2}}^{\Omega_{cc}}=\chi_{0,0}^{f11}\chi_
{\frac{3}{2},\frac{1}{2}}^{\sigma}\chi^{c}\nonumber \\
\phi_{0,-\frac{1}{2}}^{\Omega_{cc}}=&\chi_{0,0}^{f11}\chi_
{\frac{3}{2},-\frac{1}{2}}^{\sigma}\chi^{c} ~~~~~~~~~~~~
\phi_{0,\frac{3}{2}}^{\Omega_{ccc}}=\chi_{0,0}^{f12}\chi_
{\frac{3}{2},\frac{3}{2}}^{\sigma}\chi^{c}\nonumber \\
\phi_{0,\frac{1}{2}}^{\Omega_{ccc}}=&\chi_{0,0}^{f12}\chi_
{\frac{3}{2},\frac{1}{2}}^{\sigma}\chi^{c}~~~~~~~~~~~~
\phi_{0,-\frac{1}{2}}^{\Omega_{ccc}}=\chi_{0,0}^{f12}\chi_
{\frac{3}{2},-\frac{1}{2}}^{\sigma}\chi^{c} \nonumber
\end{align}

\end{document}